\documentclass[twocolumn]{aastex63}
\usepackage[caption=false]{subfig}
\usepackage{amsmath}
\usepackage{relsize}

\newcommand{\vect}[1]{\boldsymbol{#1}}

\newcommand{\e}{\text{e}}
\received{\today}
\revised{MM DD, YYYY}
\accepted{MM DD, YYYY}
\submitjournal{ApJ}

\shorttitle{Triggering diffusive tides by nonlinear mode interactions}
\shortauthors{Yu, Weinberg, Arras}
\graphicspath{{./}{figures/}}
\begin{document}

%\title{Nonlinear effects on the diffusive tidal evolution}
% \title{Diffusive Tidal Evolution including Nonlinear Mode Interactions}
\title{Tides in the high-eccentricity migration of hot Jupiters: \\
Triggering diffusive growth by nonlinear mode interactions}

\correspondingauthor{Hang Yu}
\email{hangyu@caltech.edu}

\author[0000-0002-6011-6190]{Hang Yu}
\affiliation{TAPIR, Walter Burke Institute for Theoretical Physics, Mailcode 350-17 California Institute of Technology, Pasadena, CA 91125, USA}

\author[0000-0001-9194-2084]{Nevin N. Weinberg}
\affiliation{Department of Physics, University of Texas at Arlington, Arlington, TX 76019, USA}

\author[0000-0001-5611-1349]{Phil Arras}
\affiliation{Department of Astronomy, University of Virginia, P.O. Box 400325, Charlottesville, VA 22904, USA}



\begin{abstract}
High eccentricity migration is a possible formation channel for hot Jupiters.  However, in order for it to be consistent with the observed population of planets, tides must circularize the orbits in less than $\approx$ a  Myr.   
A potential mechanism for such rapid circularization is the diffusive growth of the tidally driven planetary f-mode.  Such growth occurs if the f-mode’s phase at pericenter varies chaotically from one pericenter passage to the next.  
Previous studies focused on the variation of the orbital period due to tidal back-reaction on the orbit as the source of chaos. Here we show that nonlinear mode interactions can also be an important source. 
Specifically, we show that nonlinear interactions between a parent f-mode and daughter f-/p-modes induce an energy-dependent shift in the oscillation frequency of the parent.  This frequency shift varies randomly from orbit to orbit because the parent’s energy varies.  As a result, the parent’s phase at pericenter varies randomly, which we find can trigger it to grow diffusively.  
We show that the phase shift induced by nonlinear mode interactions in fact dominates the shift induced by tidal back-reaction and significantly lowers the one-kick energy threshold for diffusive growth by about a factor of 5 compared to the linear theory's prediction. Nonlinear interactions could thus enhance the formation rate of hot Jupiters through the high-eccentricity migration channel and potentially mitigate the discrepancy between the observed and predicted occurrence rates for close-in gas giants as compared to those further from the star. 
\end{abstract}



\section{Introduction}
\label{sec:intro}

More than 25 years after the first detection of a hot Jupiter \citep{Mayor:95}, we still do not know their dominant formation channel.
Possibilities include in situ formation, gas disk migration, and high-eccentricity tidal migration (see \citealt{Dawson:18} for a review). In the latter scenario, the planet is born beyond the snow line at $\gtrsim 1 \textrm{ AU}$ and is driven to high eccentricity through planet-planet scattering (e.g., \citealt{Rasio:96, Chatterjee:08}) or secular interactions with another planet or star (e.g., \citealt{Wu:03, Nagasawa:08, Wu:11, Hamers:17, Teyssandier:19}).  Strong tidal interactions during close pericenter passages subsequently damp the eccentricity and shrink the semi-major axis, culminating  in a planet that resides in a days-long circular orbit. 

An outstanding problem with this formation channel is the lack of very high eccentricity systems ($e>0.9$) among the observed population of hot Jupiters. In order to sufficiently speed up the circularization and thereby reduce the likelihood of catching a planet in the high-$e$ state, the tidal dissipation must be at least ten times more efficient than our own Solar System’s Jupiter \citep{Socrates:12, Dawson:15}. Such an enhanced efficiency is not necessarily inconceivable, however, as tidal dissipation in hot Jupiter systems can be sensitive to the strength and frequency of the tidal forcing and the structure of the components (see, e.g., \citealt{Ogilvie:04, Barker:11, Essick:16}). One should therefore consider the problem from first-principles rather than rely on parametrized extrapolations.  

Indeed, \citet{Wu:18} showed that the diffusive growth of the the planet's $l=2$ $f$-mode during high-eccentricity migration could lead to very rapid orbit circularization  (see also \citealt{Vick:18, Vick:19}).
This process was first considered in the hot Jupiter context by \citet{Ivanov:04} and has also been considered in a number of other high eccentricity systems; e.g., in tidal capture binaries \citep{Kochanek:92, Mardling:95} and eccentric neutron star binaries \citep{Vick:18}.  
 In \citeauthor{Wu:18}'s calculations, the $f$-mode's phase is randomly perturbed by the backreaction of the tide on the orbit, causing the mode amplitude to grow diffusively over many pericenter passages.  She argued that the $f$-mode will damp nonlinearly when its amplitude reaches unity and the mode breaks near the planet's surface.  Within $\sim 10^4\textrm{ yr}$, the planet is transported from a few AU to $\sim 0.2\textrm{ AU}$ and its eccentricity is decreased from near unity to $<0.9$. Such a rapid circularization is equivalent to a remarkably small tidal quality factor of $Q\sim 1$, five orders of magnitude smaller than Jupiter's. 
 
 \citet{Wu:18} showed that an additional feature of the diffusive growth scenario is that a planet that is secularly perturbed to high eccentricity will dynamically decouple from its perturbers when its pericenter distance reaches $\sim 4$ tidal radii. All migrating Jupiters will therefore park safely outside the zone of tidal disruption, where they are observed today.  This may explain why hot Jupiters are formed more efficiently than previous investigations of high-eccentricity tidal migration found: whereas observations show that the observed ratio of hot to cold Jupiters is $\sim 10\%$, previous theoretical calculations, which did not consider diffusive growth, yielded a ratio of only $\sim 1\%$ due to their comparatively high rates of tidal disruption (see Table 2 in \citealt{Dawson:18}).

In this paper, we extend the work of \citet{Wu:18} and \cite{Vick:18} by considering the effects of weakly nonlinear mode interactions on the $f$-mode's diffusive growth.  
We show that the random changes in mode phase induced by three-wave nonlinear interactions act in concert with tidal backreaction on the orbit in order to lower the threshold for diffusive growth. We find that for a given orbital period, the diffusive growth can be triggered at a larger pericenter distance (i.e., smaller eccentricity) and hence smaller kick amplitude, as compared to calculations including only linear physics.  Here we focus on the triggering and early phases of diffusive growth; an investigation of the long term evolution will be left to subsequent papers.

The plan of the paper is as follows. Section \ref{sec:linear} contains a review of the coupled equations for the mode and orbit including only linear processes. Following \citet{Vick:18}, an iterative map for the mode amplitude and orbital period is outlined, including the effects of planetary rotation. Nonlinear coupling of the excited f-modes with other f- and p-modes is discussed in Section \ref{sec:nonlinear}. The non-resonant phase shift and damping rate are derived, and gas giant planet models are used to evaluate the frequencies, damping rates and coupling coefficients. The nonlinear phase shift and damping are incorporated into the iterative mapping algorithm in Section \ref{sec:iter_map}. Results for short timescale simulations using the maps are presented in Section \ref{sec:trigger_growth} and conclusions and discussion are presented in Section \ref{sec:conclusions}.

\section{Linear Problem} 
\label{sec:linear}

In this section, we review the linear problem and introduce some of the notation and approximations we will use throughout our study.  In Sec.~\ref{sec:lin_prob}, we present the set of equations needed to construct an iterative map of the coupled mode-orbit evolution in  linear theory (including the Doppler shifts caused by rotation). Then in Sec.~\ref{sec:ang_momentum}, we justify our approximate treatment of angular momentum transfer and the orbital and spin evolution of the planet.

The eigenmodes form a complete basis for the fluid displacements $\vect{\xi}(\vect{x}, t)$, which can be expanded as \citep{Schenk:02}
\begin{equation}
\left\{
\begin{array}{l}
\vect{\xi}(\vect{x},t) = \sum_a q_a(t) \vect{\xi}_a(\vect{x},t) \\
\dot{\vect{\xi}}(\vect{x},t) = \sum_a (-i \omega_a) q_a(t) \vect{\xi}_a(\vect{x},t), \\
\end{array}
\right.
\end{equation}
where $\omega_a$ is the eigenfrequency of a mode and $q_a$ its amplitude. 
The sums run over both radial and angular quantum numbers as well as modes with positive and negative frequencies. 
We normalize each mode such that 
\begin{equation}
    2\omega_a^2\int d^3 x \rho \vect{\xi}^\ast_{a'}\cdot\vect{\xi}_a = E_0\delta_{aa'},
\end{equation}
where $E_0=GM^2/R$, $M$ is the mass of the planet, $R$ is its radius, and other quantities have their usual meaning. 
If we ignore nonlinear effects, the equation of motion of a planetary mode in the frame corotating with the planet is  \citep{Schenk:02}
\begin{equation}
\dot{q}_a + (i \omega_a + \gamma_a) q_a 
	= i \omega_a U_a,  \label{eq:ode_mode_amp_linear}
\end{equation}
where $\omega_a$ and $\gamma_a$ are the eigenfrequency and linear damping rate of the mode. The amplitude of
the tidal force acting on the mode is
\begin{equation}
U_a(t) =W_{lm} Q_{a} \left( \frac{M_\ast}{M} \right)  \left[\frac{R}{D(t)}\right]^{l+1} {\rm e}^{-i m \left[\Phi(t) - \Omega_{\rm s}t\right]}, 
\label{eq:U_a}
\end{equation}
where  $D$ is the orbital separation, $\Phi$ is the orbital phase, $\Omega_{\rm s}$ is the spin of the planet, the tidal overlap $Q_a=(MR^l)^{-1}\int d^3r\rho \vect{\xi}^\ast \cdot \nabla (r^l Y_{\rm lm}) $, and at leading order ($l=2$), the nonvanishing $W_{lm}$ coefficients are $W_{2\pm2}=\sqrt{3\pi/10}$ and $W_{20}=-\sqrt{\pi/5}$. 
Note that we include the Doppler shift of frequency due to rotation but ignore 
corrections to the rotating-frame frequency and eigenfunction. Also note that our sign convention is different from that used in \citet{Wu:18}. Specifically, a prograde (retrograde) mode has $m_a>0$ ($m_a<0$) in our definition.

The mode amplitudes couple to the orbital motion through the accelerations $a_D$ and $a_\phi$ in the equations of motion
\begin{eqnarray}
    &\ddot{D} &= D\dot{\Phi}^2 - \frac{G(M+M_\ast)}{D^2} + a_D,\label{eq:ddD}\\
    & D\ddot{\Phi} &= -2\dot{D}\dot{\Phi} + a_\Phi,\label{eq:D_ddPhi}
\end{eqnarray}
where to linear order
\begin{eqnarray}
    &a_D &%\simeq 
    =-\frac{E_0}{\mu D}\sum_a (l_a+1) {\rm Re}\left[q_a^\ast U_a \right], \\
    &a_\Phi &%\simeq
    =\frac{E_0}{\mu D} \sum_a m_a {\rm Im}\left[q_a^\ast U_a\right], 
\end{eqnarray}
$\mu=MM_\ast / (M+M_\ast)$ is the reduced mass.
Throughout our study, we drop the nonlinear tidal back-reaction terms as their effect on the one-kick amplitude is subdominant. 

\subsection{Iterative map including only linear effects}
\label{sec:lin_prob}

The direct integration of the coupled mode-orbit evolution equations is computationally expensive. To obtain the approximate secular evolution, an iterative mapping procedure has been developed (see, e.g., \citealt{Vick:18}), whose key steps we summarize below (see also similar derivations in \citealt{Vick:19}).

To do so, we first perform a phase shift to 
transform  Eq.~(\ref{eq:ode_mode_amp_linear}) from the
\emph{corotating} frame to the \emph{inertial} frame 
\begin{equation}
    \dot{q}'_a + \left(i \omega_a'  + \gamma_a\right) q_a' 
	= i \omega_a U_a'(t), 
	\label{eq:ode_mode_amp_inertial}
\end{equation}
where $q_a'=q_a\exp\left(-i m_a \Omega_{\rm s}t\right)$, $\omega_a'=\omega_a + m_a\Omega_{\rm s}t$, and $U_a'=U_a\exp\left(-i m_a \Omega_{\rm s}t\right)$ are, respectively, the mode amplitude, mode frequency, and tidal driving in the \emph{inertial} frame. The general solution of $q_a'$ is 
\begin{equation}
    q_a'(t) = e^{-(i\omega_a'+\gamma_a) t} \int^t i \omega_a U_a'(\tau) e^{(i\omega_a' + \gamma_a) \tau} d\tau.
\end{equation}
Suppose we know the mode amplitude right before the $k$th pericenter passage. We can then write the mode amplitude in the $k$th orbit as (see also \citealt{Vick:19})
\begin{eqnarray}
    &q_{a, k}'^{\rm (0)} &= q_{a, k-1}'^{\rm (1)} + \Delta q_{a, 1} \label{eq:iter_map_interation}\\
    &q_{a, k}'^{\rm (1)} &= q_{a, k}'^{\rm (0)} e^{-(i\omega_a'+\gamma_a) P_{{\rm orb}, k}},
    \label{eq:iter_map_mode_prop}
\end{eqnarray}
where the superscript $(0)$ and $(1)$ indicate that the amplitudes are respectively evaluated right after and right before a pericenter passage. The quantity $\Delta q_{a, 1}$ is the one-kick amplitude the mode receives at the pericenter. It is given by\footnote{Formally the integration should be performed from right before the $k$'th pericenter passage to right before the next passage. 
We can nonetheless shift the initial time because $U_a'(t)=U_a'(t+\sum_{k'} P_{{\rm orb}, k'})$ in the inertial frame if the pericenter distance stays approximately fixed throughout the evolution. This also is the reason the one-kick amplitude can be treated as a constant for different orbital cycles. 
}
\begin{eqnarray}
    \Delta q_{a, 1} =&
    {\mathlarger \int} &
    i\omega_a U'_a(\tau) e^{(i\omega'_a+\gamma_a)\tau} d\tau.
    \label{eq:one_kick_amp}
\end{eqnarray}
In the equation above, the integration is preformed over one orbital period. 
Since we care about orbits that are highly eccentric, we make the approximation that the tidal interaction happens only near pericenter. Therefore, the limits of integration in Eq.~(\ref{eq:one_kick_amp}) can be dropped as long as they bracket the pericenter passage. 

It is convenient to define an orbital integral $K_{lm}$ as\footnote{Note that a tidal field with spherical degree $(l, m)$ linearly couples to a mode with $l_a=l$ and $m_a=m$ due to the angular integral in $Q_a$. 
}~\citep{Press:77} 
\begin{equation}
    K_{lm}(\omega) = \frac{\omega_0 W_{lm}}{2\pi}\int  \left[\frac{D_{\rm peri}}{D(\tau)}\right]^{l+1} e^{i\left[\omega \tau - m\Phi(\tau)\right]} d\tau,
    \label{eq:K_lm_def}
\end{equation}
where $D_{\rm peri}\equiv a_{\rm orb}(1-e_{\rm orb})$ is the pericenter distance, $e_{\rm orb}$ the eccentricity, and $\omega_0\equiv\sqrt{GM/R^3}$.  
If we ignore the perturbations on $D$ and $\Phi$, \citet{Lai:97} provide an analytical expression for $K_{22}$ (i.e., $l=m=2$) assuming the orbit is parabolic, 
\begin{eqnarray}
    K_{22}(\omega, \Omega_{\rm peri}) &\simeq& \frac{2z^{3/2}e^{-2z/3}}{\sqrt{15}}\left(\frac{\omega_0}{\Omega_{\rm peri}}\right)\left(1-\frac{\sqrt{\pi}}{4\sqrt{z}}\right), \nonumber \\
    &\simeq& 1.1\times10^{-2} \left(\frac{z}{11}\right)^{-5.7}
    \left(\frac{\omega_0}{\Omega_{\rm peri}}\right),
    \label{eq:K_lm_para}
\end{eqnarray}
where $z\equiv\sqrt{2}\omega/\Omega_{\rm peri}$ and $\Omega_{\rm peri}^2 \equiv G(M+M_\ast)/D_{\rm peri}^3$, and in the second line we have expanded the expression around $z=11$ to emphasize the steep decline.

In terms of $K_{lm}$, we can write the one-kick amplitude  as 
\begin{eqnarray}
    \Delta q_{a, 1} &=&  i 2\pi Q_a K_{lm}(\omega_a', \Omega_{\rm peri})  \nonumber \\
    &\times& \left(\frac{\omega_a\Omega_{\rm peri}^2}{\omega_0^3}\right) \left(\frac{M_\ast}{M+M_\ast}\right)\left(\frac{R}{D_{\rm peri}}\right)^{l-2}. 
    \label{eq:one_kick_amp_vs_Klm}
\end{eqnarray}
The damping term entering $\Delta q_{a,1}$ can be safely dropped because $\gamma_a \ll \Omega_{\rm peri}$. 
Note that for a parabolic orbit, $K_{lm}$ is a real number and therefore $\Delta q_{a,1}$ is purely imaginary. Also note that when calculating $K_{lm}$ one should use the mode frequency in the \emph{inertial frame} $\omega_a'=(\omega_a + m \Omega_{\rm s})$. Combining with the expansion given in Eq.~(\ref{eq:K_lm_para}), one sees immediately that the spin reduces the one-kick amplitude for a prograde mode with $m>0$ in our convention.

To account for the tidal back reaction on the orbit, we adopt an energy conservation argument instead of explicitly coupling the mode amplitude equation and the tidal accelerations $a_D$ and $a_{\Phi}$. Upon receiving a kick at  pericenter, the energy stored in a stellar mode changes by (including contributions from the mode and its complex conjugate)
\begin{equation}
    \Delta E_{a,k} = \left[|{q'}_{a, k}^{(0)}|^2 - |{q'}_{a, k-1}^{(1)}|^2\right]E_0. 
\end{equation}
Since the energy stored in the tidal coupling (the term $\propto {\rm Re}\left[q_a^\ast U_a\right]$) is small everywhere except for at the pericenter and the spin rate of the planet should stay approximately fixed (which we will justify shortly), the change in the energy of stellar modes needs to be balanced by the orbital energy, 
\begin{equation}
    \Delta E_{{\rm orb},k} = E_{{\rm orb},k} - E_{{\rm orb}, k-1} = -\Delta E_{a, k},
    \label{eq:iter_map_E_orb}
\end{equation}
where $E_{{\rm orb},k} = -GMM_\ast/2a_{{\rm orb},k}$ is the orbital energy at the $k$'th orbit.  

A direct consequence is that the change in the orbital energy also alters the orbital period $P_{\rm orb}=2\pi/\Omega_{\rm orb} = 2\pi\sqrt{a_{\rm orb}^3/G(M+M_\ast)} \propto (-E_{\rm orb})^{-3/2}$, as 
\begin{equation}
    \frac{\Delta P_{{\rm orb},k}}{P_{{\rm orb},k}} = \frac{3}{2}\frac{\Delta E_{a,k}}{E_{{\rm orb},k}}. 
    \label{eq:iter_map_P_orb}
\end{equation}
Since the value of $\Delta E_{a,k}$ is different from orbit to orbit, the orbital period varies. This, in turn, leads to a stochastic evolution of the mode's phase per orbital cycle 
\begin{equation}
    \Delta \phi_{{\rm br},k} = -\omega_a \Delta P_{{\rm orb},k} = -\frac{3}{2}\omega_a P_{{\rm orb},k} \frac{\Delta E_{a,k}}{E_{{\rm orb},k}},
    \label{eq:dphi_br_k}
\end{equation}
where we have used a subscript ``br'' to stand for the fact that this phase is due to the back reaction of the tide. As shown in previous studies~\citep{Vick:18, Wu:18}, the randomness of the phase shift $\Delta \phi_{{\rm br},k}$ is key to triggering the diffusive growth of a tidally driven mode. 

In order to simplify the notation, we will sometime omit the subscript ``$k$'' when we do not need the quantity to be evaluated at a specific orbit cycle. 

\subsection{Orbital and spin angular momentum}
\label{sec:ang_momentum}
An energy transfer is typically associated with an angular momentum transfer as well. Nonetheless, since the change in the orbital angular momentum  $\Delta L_{\rm orb}\simeq \Delta E_{\rm orb}/\Omega_{\rm peri}$, we have that at high eccentricity ~\citep{Vick:18}
\begin{eqnarray}
      \Big{|}\frac{\Delta L_{\rm orb}}{L_{\rm orb}} \Big{|}
      &\simeq& 
      \Big{|}\frac{\Delta E_a(1-e_{\rm orb})}{2\sqrt{2}E_{\rm orb}}\Big{|}
      \ll  \Big{|}\frac{\Delta E_a}{E_{\rm orb}}\Big{|}.  
      \label{eq:dlogL}
\end{eqnarray}
Therefore, the orbital angular momentum stays as a constant throughout the evolution to a very good approximation. Suppose, for example, the initial orbit is  $a_{\rm orb}=1\,{\rm AU}$ and $e_{\rm orb}=0.98$ and it evolves to $a_{\rm orb}=0.2\,{\rm AU}$ (with $e_{\rm orb}\simeq 0.9$) due to tidal dissipation. Whereas the orbital energy changes by a factor of 5, the orbital angular momentum changes by only $3\%$ in the process. Furthermore, since the angular momentum is nearly constant and 
\begin{eqnarray}
    L_{\rm orb} &=& \mu \sqrt{G(M+M_\ast)a_{\rm orb}(1-e_{\rm orb}^2)}\nonumber \\
    &\simeq& \mu\sqrt{2G(M+M_\ast)D_{\rm peri}}, 
\end{eqnarray}
at high-eccentricity, it follows that the pericenter distance $D_{\rm peri}$ is also nearly constant throughout the orbital evolution.

Moreover, under the high-eccentricity limit, the linear one-kick amplitude $\Delta q_{a,1}$ depends on the Keplerian elements only through the pericenter distance $D_{\rm peri}$ [which determines $\Omega_{\rm peri}$ for fixed $(M, M_\ast)$]. As the pericenter distance stays nearly unchanged, the one-kick amplitude $\Delta q_{a,1}$ also remains approximately  constant.

So far we have left the spin of the planet $\Omega_{\rm s}$ as a free parameter. One plausible scenario is that the planet reaches pseudo-synchronization with the orbit via the equilibrium tide, 
leading to~\citep{Hut:81}
\begin{eqnarray}
    \Omega_{\rm s}&\simeq& \frac{1+\frac{15}{2}e_{\rm orb}^2 + \frac{45}{8}e_{\rm orb}^4 + \frac{5}{16}e_{\rm orb}^6}{(1+3e_{\rm orb}^2 + \frac{3}{8}e_{\rm orb}^4)(1-e_{\rm orb}^2)^{3/2}} \Omega_{\rm orb},    \nonumber\\
    &\simeq&1.17\Omega_{\rm peri} \quad\quad (e_{\rm orb}\to 1),
    \label{eq:omega_s_ps}
\end{eqnarray}
where the second line applies in the high-eccentricity limit. A constant pericenter distance (hence constant $\Omega_{\rm peri}$) would then imply that the spin frequency as set by the pseudo-synchronization condition also stays approximately constant. 
We note that including a pseudo-synchronous rotation of the planet will increase the prograde f-mode frequency in the inertial frame, which will tend to decrease the one-kick amplitude and slow the orbital evolution as compared to the non-rotating planet case.
Nevertheless, the one-kick amplitude of the prograde mode is still two orders of magnitude greater then the $m_a=0$ mode and even more for the retrograde mode. Therefore, we will only consider the prograde mode in the subsequent discussion.

To summarize our proceedure, we discard the evolution of the angular momenta and, self-consistently, treat $\Omega_{\rm s}$ and $D_{\rm peri}$ as approximate constants during the circularization process (at least for the initial phase when $1-e_{\rm orb}\ll 1$ is well satisfied). The evolution trajectories will thus reduce to the ones studied by \citet{Vick:18} and \citet{Wu:18} as long as one uses the mode frequency in the inertial-frame $\omega_a'=\omega_a + m_a \Omega_{\rm s}$ and neglects the nonlinear effects described in the next section.

\section{Nonlinear Problem}
\label{sec:nonlinear}
We now consider how weakly nonlinear effects modify the problem.  At lowest nonlinear order, the amplitude equation of a mode $a$ is~\citep{Weinberg:12}
\begin{equation}
\dot{q}_a + (i \omega_a + \gamma_a) q_a 
	= i \omega_a \left[U_a + \sum_b U_{ab}^\ast q_{b}^\ast + \sum_{bc}\kappa_{a b c} q_b^\ast q_c^\ast \right], \label{eq:ode_mode_amp_general}
\end{equation}
where $U_{ab}$ is the nonlinear tide, $\kappa_{abc}$ is the three-mode coupling coefficient, and asterisks denote complex conjugation. There is a significant cancellation between the nonlinear tide and three-mode coupling to the equilibrium tide such that $U_{ab}+2\sum_c \kappa_{abc} U_c\simeq0$ \citep{Weinberg:12}.  By treating the cancelation as perfect, we have
\begin{eqnarray}
    \dot{q}_a + (i \omega_a &+& \gamma_a) q_a 
	= i \omega_a U_a  \nonumber \\
	&+& i\omega_a \sum_{bc}\kappa_{abc} \left( q_b^\ast q_c^\ast -q_b^\ast U_c^\ast - q_c^\ast U_b^\ast \right). \label{eq:ode_mode_amp_no_Uab}    
\end{eqnarray}
We will solve this equation (or approximate its solution), in order to determine how nonlinear mode interactions influence the diffusive growth of the $f$-mode and thereby a planet's high-eccentricity tidal migration.

Since the daughters' direct, linear coupling to the tide is small, we  expect the most significant nonlinear effect to be the modification of the parent mode's free evolution away from pericenter (when $U_a\simeq 0$). Specifically, we show in Section~\ref{sec:nl_freq_damp} that the nonlinear mode couplings can be viewed as energy-dependent shifts of the parent's eigenfrequency and damping rate. We then derive the time-dependent evolution of such a nonlinear oscillator in Section~\ref{sec:prop_nl_oscillator}. Lastly, we examine the nonlinear effects in a typical Jupiter model in Section~\ref{sec:values_for_real_jupiters}.

\subsection{Nonlinear frequency shift and effective damping rate}
\label{sec:nl_freq_damp}
Previous studies have shown that at  linear order, the $l_a=m_a=2$ f-mode with $\omega_a>0$ has the greatest energy and that it dominates the orbital evolution (see, e.g., \citealt{Wu:18}).
We will thus focus on a single parent mode (mode $a$) with $l_a=m_a=2$ and $\omega_a>0$ (and its complex conjugate to get real, physical quantities). We consider the nonlinear effects due to this parent mode coupling to itself and  a daughter mode (which can be another f-mode or a p-mode), which results in the inhomogeneous driving of the daughter.\footnote{As we consider fully convective Jupiter models, there are no low-frequency g-modes that can parametrically couple to the parent~\citep{Weinberg:12}. Therefore, the non-resonant nonlinearity considered in this work should be distinguished from the parametric instabilities considered in, e.g., \citet{Essick:16} for solar-type stars and \citet{Yu:20a} for white dwarfs.}  

Once we know the parent mode's angular pattern $(l_a, m_a)$, we can further utilize the three-mode angular selection rules and divide up the nonlinear couplings into two categories, which we will refer to as $aab$ and $aa^\ast c$, respectively. 

In the $aab$ case, the driving is formed by mode $a$ coupling to itself. By the angular section rule, only daughter modes with $l_b  = -m_b = 4$ can couple to this driving. We will refer to such a daughter as mode $b$ and note that it is forced at a frequency $-2\omega_a$. 

By contrast, in the $aa^\ast c$ case, mode $a$ couples to its complex conjugate $a^\ast$ and drives a daughter mode (mode $c$) with $m_c=0$ and $l_c=0,2,4$. In this case, mode $c$ experiences a forcing at zero frequency.\footnote{Care must be taken when such DC forcing is encountered in the nonlinear problem in order to ensure that the forcing represents the physical transfer of energy and angular momentum between distinct oscillation modes, and not just a constant nonlinear modification of the linear mode frequency and eigenfunction. 
In the present situation, there is additional time-dependence due to the mode amplitudes, which change from one orbit to the next, and such forcing in turn  leads to further time-dependent changes in the modes' amplitude and phase.\label{footnote:DC_forcing}}

To make the abstract problem more transparent, we write out the explicit three-mode coupling equations for both cases. For simplicity, we start by considering only a single mode $b$ and a single mode $c$. We will perform a summation over modes in the end to obtain the general solution. We will also drop  the couplings involving more than one daughter mode for analytical simplicity; all the allowed couplings are included in the numerical calculations when we validate our analytical approximations. We then have
\begin{eqnarray}
    &\dot{q}_a+(i\omega_a+\gamma_a)q_a &= 2i \omega_a \kappa_b q_b^\ast q_a^\ast + 2i\omega_a\kappa_c q_c^\ast q_a,\label{eq:dqa_dt}\\
    &\dot{q}_b+(i\omega_b+\gamma_b)q_b &= i \omega_b \kappa_b q_a^\ast q_a^\ast,\label{eq:dqb_dt}\\
    &\dot{q}_c+(i\omega_c+\gamma_c)q_c &= 2i \omega_c \kappa_c q_a^\ast q_a.\label{eq:dqc_dt}
\end{eqnarray}
Note that the above set of equations describes the evolution away from  pericenter, after the parent has received its most recent kick,   since here we are interested in following the parent's free (i.e., unforced) evolution leading up to the next pericenter passage. Consequently, we do not include any tidal forcing terms. 

To seek the leading-order nonlinear correction, we solve Eqs.~(\ref{eq:dqa_dt})-(\ref{eq:dqc_dt}) in a perturbative manner. Away from pericenter and without nonlinear couplings, we have $q_a\sim \exp(-i\omega_a t)$. Using this as the driving term, the steady-state solutions of the daughters are\footnote{These solutions involve some approximations that we discuss in Section~\ref{sec:iter_map} and Appendix~\ref{sec:q_b}.} 
\begin{eqnarray}
    & q_b &= \frac{\left.\omega_b(2\omega_a + \omega_b) + i \omega_b\gamma_b\right.}{(2\omega_a + \omega_b)^2 + \gamma_b^2}\kappa_b q_a^\ast q_a^\ast, \label{eq:q_b_ss_soln}\\
    & q_c &= \frac{2\omega_c^2+2i\omega_c\gamma_c}{\omega_c^2 + \gamma_c^2}2\kappa_c q_a^\ast q_a. \label{eq:q_c_ss_soln}
\end{eqnarray}
Plugging the daughter modes above back in to Eq.~(\ref{eq:dqa_dt}), we obtain
\begin{equation}
    \dot{q}_a + \left[i\left(\omega_a + \delta \omega_a\right) + \left(\gamma_a + \delta \gamma_a\right)\right] q_a = 0,
\end{equation}
where\footnote{We use ``$\Delta$'' to indicate the difference between adjacent orbital cycles, and ``$\delta$'' for the nonlinear deviation relative to the linear case. } 
\begin{eqnarray}
    &\delta \omega_a = -\omega_a&\left[\frac{2\omega_b(2\omega_a+\omega_b)}{(2\omega_a+\omega_b)^2+ \gamma_b^2} \kappa_b^2 \right. \nonumber \\
    &&\left.  + \frac{4\omega_c^2}{\omega_c^2 + \gamma_c^2}\kappa_c^2\right]\tilde{E}_a \\
    &\delta \gamma_a = 
    -\omega_a & \left[\frac{2\omega_b\gamma_b}{(2\omega_a+\omega_b)^2+\gamma_b^2} \kappa_b^2 \right. \nonumber \\
    &&\left. + \frac{4\omega_c \gamma_c}{\omega_c^2 + \gamma_c^2}\kappa_c^2 \right]\tilde{E}_a,
\end{eqnarray}
and $\tilde{E}_a\equiv q_a^\ast q_a$ is the dimensionless energy of mode $a$.\footnote{For the rest of the paper, we will use the ``tilde'' symbol to represent dimensionless energies (i.e., energies normalized by the natural energy of the planet, $E_0=GM^2/R$).} The physical mode energy including the contribution from both $a$ and its complex conjugate $a^\ast$ is $E_a=\tilde{E}_a E_0$ in our normalization, where $E_0=GM^2/R$ is the natural energy of the planet.  

We thus see that the leading-order nonlinear correction corresponds to a shift in the eigenfrequency (conservative part) of the parent mode and an excess damping term (dissipative part), both of which depend linearly on the energy of the parent mode (see also \citealt{Landau:76, Kumar:94, Kumar:96}). We can therefore define
\begin{eqnarray}
    &\delta \omega_a (\tilde{E}_a) &= \Omega \tilde{E}_a,\label{eq:domega_leading_order}\\
    &\delta \gamma_a (\tilde{E}_a) &= \Gamma \tilde{E}_a, \label{eq:dgamma_leading_order}
\end{eqnarray}
where, after putting back the summation over all the daughters that couple to mode $a$, we have
\begin{eqnarray}
    &\Omega = -\omega_a&\left[\sum_b \frac{2\omega_b(2\omega_a+\omega_b)}{(2\omega_a + \omega_b)^2 + \gamma_b^2}\kappa_b^2 \right. \nonumber \\
    &&\left. + \sum_c \frac{4\omega_c^2}{\omega_c^2 + \gamma_c^2}\kappa_c^2\right], \label{eq:Omega}\\
    &\Gamma = \sum_b &\frac{-2\omega_a\omega_b}{(2\omega_a+\omega_b)^2+\gamma_b^2}\gamma_b \kappa_b^2.\label{eq:Gamma}
\end{eqnarray}

Note that mode $c$ does not contribute to the nonlinear damping $\Gamma$. Mathematically, this can be understood by noticing that for every mode $c$ with $\omega_c$, there exists a mode $c^\ast$ with $-\omega_c$ (i.e., the complex conjugate of $c$; they both have $m_c=m_{c^\ast}=0$) that has the exact opposite contribution to $\Gamma$. Therefore, after summing over the $\pm \omega_c$ pair, the nonlinear dissipation due to mode $c$ cancels exactly. Physically, we can view mode $c$ as a nonlinear modification of the planet's structure, which changes the frequency at which the parent wave propagates (see footnote \ref{footnote:DC_forcing}). Nonetheless, mode $c$ is not a wave itself (as it is non-oscillatory) and therefore it does not contribute to the energy dissipation. 

By contrast, mode $b$ (corresponding to a nonlinearly excited wave oscillating at $2\omega_a$) enhances the dissipation, as one would expect physically. After the summation, $\delta \gamma_a$ is always positive because a mode $b$ with negative (positive) frequency would contribute a positive (negative) dissipation rate, and $(2\omega_a + |\omega_b|) > (2\omega_a - |\omega_b|)$ as the parent mode has $\omega_a>0$. Consequently, we obtain a net increase in the damping after summing over each $\pm \omega_b$ pair. 

Now turn to the nonlinear frequency shift $\Omega$. Its sign is not definite. While most of the modes\footnote{This includes all of the mode $c$ type modes (with $m_c=0$), all the positive-frequency mode $b$ type modes (with $m_b=-4$ and $\omega_b>0$), and the negative-frequency ones with $|\omega_b| > 2\omega_a$. 
} act to  reduce the parent mode's frequency, a mode $b$ with $\omega_b<0$ and $(2\omega_a + \omega_b)>0$ 
will increase the parent mode's frequency. In practice, we find that only the $l_b=-m_b=4$ f-mode satisfies the condition $\omega_b(2\omega_a + \omega_b)<0$; this mode can in fact be resonant with the parent, although we find that it is only a mild resonance since  the mode spectrum is sparse for low-order (in both $n$ and $l$) modes. Therefore, in general we would expect $\Omega<0$. However, in principle it could be positive if there is a rare strong resonance such that $(2\omega_a + \omega_b)\simeq 0^{+}$.\footnote{Note that the detuning $(2\omega_a + \omega_b)$ is not affected by the choice of reference frame (inertial or corotating) as one would expect physically. This is guaranteed by the angular selection rule $2m_a+m_b=0$, which exactly cancels  the Doppler shifts.} 

\subsection{Evolution of the nonlinear oscillator}
\label{sec:prop_nl_oscillator}

In the previous section, we showed that nonlinear mode interactions perturb the frequency of the $f$-mode by $\delta\omega_a(\tilde{E}_a)$ and its damping rate by $\delta\gamma_a (\tilde{E}_a)$. These cause a dephasing of the $f$-mode $\delta\phi_{\rm nl}= -\int \delta\omega_a dt$ in excess of the back reaction of the $f$-mode on the tide considered in previous studies of diffusive growth (e.g., \citealt{Wu:18}). Furthermore, this change in phase varies from orbit to orbit due to the changes in the parent energy. Nonlinear effects can therefore contribute to, and even trigger (as we will show), diffusive growth.  Since $\delta\omega_a$ depends on the energy of the mode $\tilde{E}_a$, in order to determine $\delta\phi_{\rm nl}$ as a function of time,  we need to determine how $\tilde{E}_a$ evolves. Note that here we focus on the evolution when the planet is far from pericenter, i.e., of the free oscillator. 
The goal of this section is therefore to determine $\tilde{E}_a(t)$ over an orbit, 
and from it calculate the nonlinear contributions to the dephasing $\delta\phi_{\rm nl}$. The construction of an iterative mapping from orbit to orbit similar to that of the linear studies will be discussed later in Section~\ref{sec:iter_map}. 

The energy evolution is given by\footnote{Here we have implicitly assumed that the parent mode's energy dominates the total energy stored in the stellar oscillations, which is a reasonable approximation in the case we consider here.} 
\begin{equation}
    \dot{\tilde{E}}_a + 2\left[\gamma_a + \delta \gamma_a(\tilde{E}_a)\right] \tilde{E}_a = 0.
    \label{eq:dE_a/dt}
\end{equation}
If we substitute in Eq.~(\ref{eq:dgamma_leading_order}) for $\delta \gamma_a (\tilde{E}_a)$, then the above equation can be solved easily as 
\begin{eqnarray}
    &\tilde{E}_a(t) &= \frac{\gamma_a \tilde{E}_a^{(0)}}{-\Gamma \tilde{E}_a^{(0)} + \left[\gamma_a + \Gamma \tilde{E}_a^{(0)}\right] e^{2\gamma_a t}} \nonumber \\
    &&\simeq \frac{\tilde{E}_a^{(0)}}{1 + 2\left[\gamma_a + \Gamma \tilde{E}_a^{(0)}\right]t},\label{eq:E_vs_t}
\end{eqnarray}
where $\tilde{E}_a^{(0)}$ is the initial mode energy and in the second line we expand $\exp\left[2\gamma_a t\right]\simeq (1 + 2\gamma_a t)$. This is a good approximation because the linear damping of the parent mode is typically small  ($1/\gamma_a\simeq 10^9\,{\rm yr}$ for the Jupiter model we consider) and over the course of a $\sim 1\,{\rm yr}$ orbit, the condition $\gamma_a t\ll 1$ is very well satisfied. 

The total accumulated phase can be written as $\phi_a(t) = -\int [\omega_a' + \delta \omega_a(\tilde{E}_a)] dt$. Of particular interest is the excess dephasing due to nonlinear interactions
\begin{eqnarray}
    \delta \phi_{\rm nl} 
    = -\int \delta \omega_a [\tilde{E}_a(t)] dt  
    = -\int  \frac{\delta \omega_a (\tilde{E}_a)}{d\tilde{E}_a/dt} d \tilde{E}_a, 
\end{eqnarray}
where we use the subscript ``nl'' to indicate the excess phase due to nonlinear effects (in contrast to tidal back reactions denoted by a subscript ``br''), and in the second equality we change variables from  time $t$ to energy $\tilde{E}_a$. If we use Eq.~(\ref{eq:domega_leading_order}) for $\delta \omega_a$ and Eq.~(\ref{eq:dE_a/dt}) for $d\tilde{E}_a/dt$, then the nonlinear dephasing is
\begin{equation}
    \delta{\phi_{\rm nl}}(\tilde{E}_a) = \frac{\Omega}{2\Gamma}\ln \left[\frac{\gamma_a + \Gamma\tilde{E}_a}{\gamma_a + \Gamma\tilde{E}^{(0)}}\right].
    \label{eq:dphi_vs_E_leading_order}
\end{equation}
We can also write the (leading-order) dephasing as a function of time by plugging in Eq.~(\ref{eq:E_vs_t}). In the limit that the parent mode's dissipation is small ($\gamma_a t\to 0$), we can cast the dephasing in an intuitive form as
\begin{eqnarray}
    \delta{\phi_{\rm nl}}(t)&\simeq& -\frac{\Omega}{2\Gamma} \ln \left[1+2\Gamma \tilde{E}_a^{(0)} t\right], \label{eq:dphi_vs_t_leading_order}\\
    &\simeq& -\Omega \tilde{E}_a^{(0)} t, \label{eq:dphi_vs_t_leading_order_small_diss}
\end{eqnarray}
where in the second equality we further assumed $2\Gamma_a \tilde{E}_a^{(0)}t\ll 1$. For $\tilde{E}_a^{(0)}\simeq 10^{-3}$ and $t\simeq 1\,{\rm yr}$, this condition is satisfied if $\Gamma/\omega_a < 10^{-5}$. As we will see shortly in the following section and Table~\ref{tab:models}, Eq.~(\ref{eq:dphi_vs_t_leading_order_small_diss}) is well satisfied for the Jupiter model we consider in this work as it has a weak damping. Nonetheless, for Jupiters with greater radii, the damping rate can be significantly higher~\citep{Arras:09} and Eq.~(\ref{eq:dphi_vs_t_leading_order}) should be used instead.
Obviously, Eq.~(\ref{eq:dphi_vs_t_leading_order_small_diss}) also applies for conservative systems.

\subsection{Values of $\Omega$ and $\Gamma$ for a Jupiter model}
\label{sec:values_for_real_jupiters}

From the discussion above, we see that the leading-order nonlinear corrections to the parent mode correspond to shifts in both the eigenfrequency and the damping rate that are linearly proportional to the mode energy $\tilde{E}_a$ [Eqs.~(\ref{eq:domega_leading_order}) and (\ref{eq:dgamma_leading_order})]. The nonlinear dynamics can thus be characterized by the two coefficients $\Omega$ and $\Gamma$ (both having the dimension of frequency in our definition). These coefficients further depend on the parent mode's eigenfrequency (with $\Omega, \Gamma\propto \omega_a$) and the properties of the daughters. We now describe values for $\Omega$ and $\Gamma$ for a Jupiter model with $M=M_{\rm J}$ and $R=1.1\,R_{\rm J}$. 

Using \texttt{MESA} (version 10398; \citealt{Paxton:11, Paxton:13, Paxton:15, Paxton:18, Paxton:19}), we construct a planetary model with a total mass equal to Jupiter's $M=M_{\rm J}$ and a core mass of $5\,M_\oplus$ (though the results should be insensitive to the core as the eigenfunctions of both f- and p-modes are largest near the surface). 
We then let the model contract until it reaches a desired radius, which we choose to be $R=1.1R_{\rm J}$. Irradiation is turned on in this contraction phase with a fixed flux of $6.8\times10^{6}\,{\rm erg\, cm^{-2}\, s^{-1}}$, which corresponds to the average flux the planet receives assuming an orbit with $(a_{\rm orb}, e_{\rm orb})=(1\,{\rm AU}, 0.98)$ and a solar-type host star (the equilibrium temperature is $\simeq 420\,{\rm K}$).  
After the construction of the background model, we find the (adiabatic) eigenmodes (including both the parent f-mode and the leading-order daughter f- and p-modes) using \texttt{GYRE}~\citep{Townsend:13, Townsend:18}. The three-mode coupling coefficient is calculated using the expression in \citet{Weinberg:12}. We account for the damping of each mode due to turbulent convection using the approach described in Appendix B3 of \citet{Burkart:13}. 

\begin{table}
\centering
\caption{Properties of the Jupiter model considered in our study. We write $E_0 {=} GM^2/R {=} E_{0, 43} \times 10^{43}\,{\rm erg}$, $\omega_0{=}\sqrt{GM/R^3}{=}\omega_{0,-4}\times 10^{-4}\,{\rm  rad\,s^{-1}}$
as well as $\Gamma = \Gamma_{-10} \times 10^{-10} $ 
We use the subscript ``prnt'' to denote the parent mode (the prograde f-mode with $l_a=2$) that directly couples to the tide. 
}
\begin{tabular}{ ccccccc } 
 $R$ & $E_{0,43}$ & $\omega_{0, -4}$ & $\omega_{\rm prnt}$ & $Q_{\rm prnt}$ & $\Omega$ & $\Gamma_{-10}$ \\
 \hline
 $1.1R_{\rm J}$ & $3.1$ & $5.1$ & $1.1\omega_0$ & 0.43 & $-29\,\omega_{\rm prnt}$ & $8.1\,\omega_{\rm prnt}$ \\
\end{tabular}
\label{tab:models}
\end{table}

The key parameters of the Jupiter model are summarized in Table~\ref{tab:models}. Of particular interest are the values of $\Omega$ and $\Gamma$. We find that $\Omega$ is typically negative and of order $|\Omega/\omega_{\rm prnt}|\sim 30$, where here we will use subscript ``$a$'' to stand for a generic mode and ``prnt" to indicate the parent mode specifically. The value of $\Gamma$ will always be positive, as argued in Sec.~\ref{sec:nl_freq_damp}.

\begin{figure}
   \centering
   \includegraphics[width=0.45\textwidth]{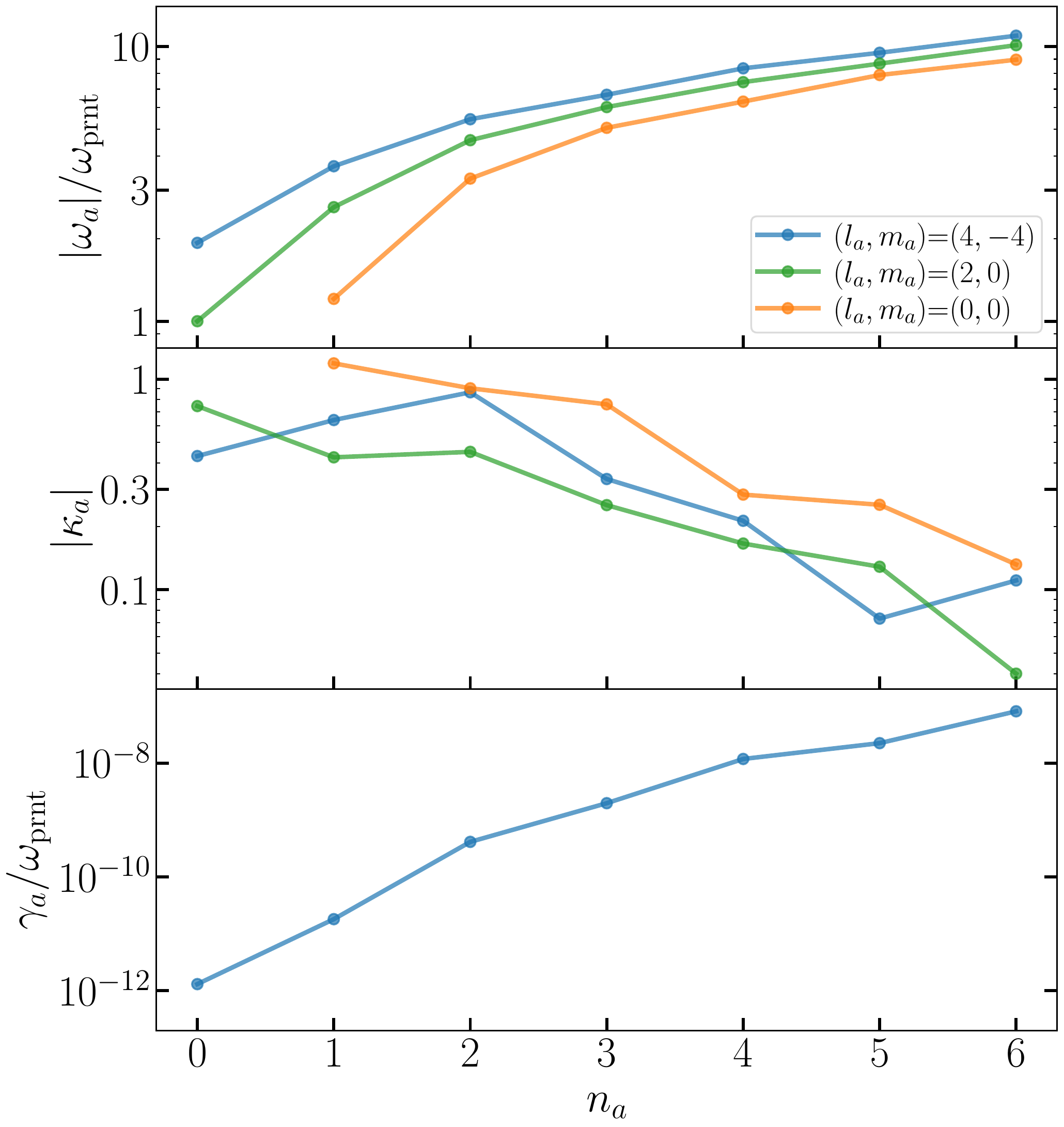} % requires the graphicx package
\caption{Eigenfrequency (top), three-mode coupling coefficient (middle),  and linear damping rate (bottom) of each daughter mode. 
Here the subscript $a$ in the labels stands for a generic mode, and we specifically label the parent mode's eigenfrequency as $\omega_{\rm prnt}$. For a daughter with $|m_a|=4$, the coupling we consider is specifically due to parent-parent-daughter, and for an $m_a=0$ daughter, it is due to parent-parent$^\ast$-daughter. 
We only show the damping rate for the $|m_a|=4$ modes as only they contribute to the energy dissipation.}
\label{fig:lin_omega_gam}
\end{figure}

\begin{figure}
   \centering
   \includegraphics[width=0.45\textwidth]{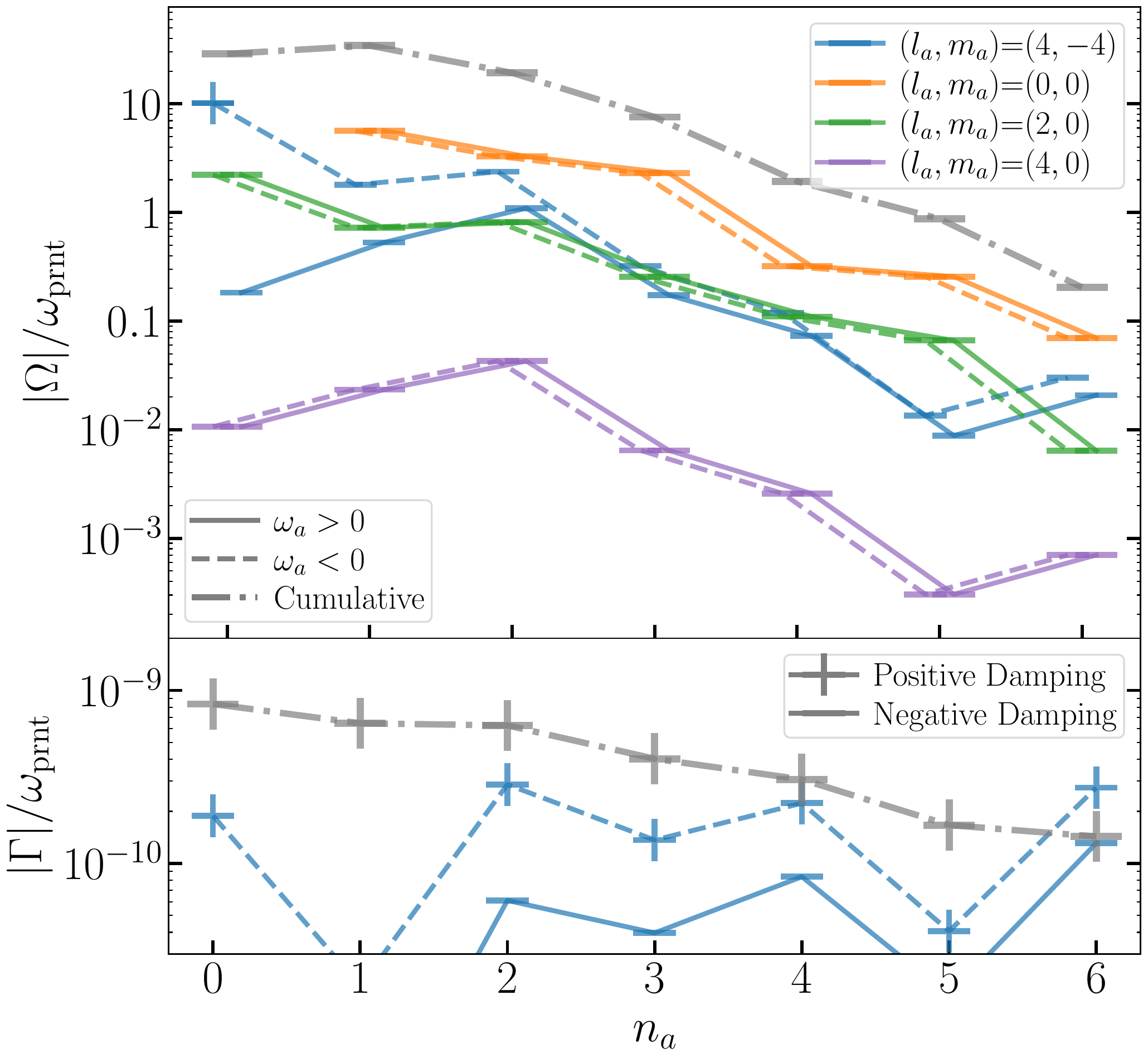} % requires the graphicx package
   \caption{
   Each daughter mode's contribution to the sums that comprise the nonlinear frequency shift $\Omega$ (top panel) and nonlinear damping $\Gamma$ (bottom panel). The x-axis is the radial order of the daughter, with $n_a=0$ for f-modes and $n_a>0$ for p-modes. We use different colors to label daughter modes with different $(l, m)$ and solid/dashed lines to label positive/negative-frequency modes. For clarity, we shifted the radial order by $+0.1$ ($-0.1$) for modes with positive (negative) frequencies. If a daughter mode contributes a positive value to $\Omega$ ($\Gamma$), a ``+'' marker is used; otherwise, we use ``-'' to show the negative contribution. In the grey, dash-dotted traces, we also show the cumulative values of $\Omega$ and $\Gamma$ obtained by summing over daughter modes with all the possible $(l_a, m_a, \omega_a)$ and with radial order $n_a'\geq n_a$. The left-most grey markers thus correspond to the values of $\Omega$ and $\Gamma$. 
    } 
   \label{fig:Omega_Gam_1p0Mj_1p1Rj}
\end{figure}

To illustrate the values of the parameters that enter the calculation of $\Omega$ and $\Gamma$,  we present the eigenfrequency, three-mode-coupling coefficients,  and the linear damping rate of each daughter in Figure~\ref{fig:lin_omega_gam}. In addition,  we  show in Figure~\ref{fig:Omega_Gam_1p0Mj_1p1Rj} each daughter's contribution to $\Omega$ (top panel) and $\Gamma$ (bottom panel). 
Note that only the negative-frequency, $(l_a, m_a, n_a)=(4, -4, 0)$ f-mode contributes a positive value to $\Omega$. Although it has the greatest single-mode contribution as it is the most-resonant daughter with respect to the parent's driving, the resonance is not particularly strong.\footnote{The Jupiter model we consider here has $|\omega_b/(2\omega_a + \omega_b)|\simeq 27$. As a comparison, for a constant density, incompressible sphere, the f-mode eigenfrequency follows $\omega_a {\propto} \sqrt{2l_a(l_a-1)/(2 l_a + 1)}$, which leads to a similarly large number $|\omega_b/(2\omega_a + \omega_b)|\simeq 10$. } Instead, upon summing over all couplings, the value of $\Omega$ is dominated by the coupling to the many $m_a=0$ modes (``mode c'' in Sec.~\ref{sec:nl_freq_damp}) as well as the off-resonant $m_a=-4$ modes (``mode b'' with $|\omega_b|> 2|\omega_a|$) and is therefore \emph{negative}.

By contrast, only the $m_a=-4$ daughters (oscillating at $-2\omega_{\rm prnt}$) contribute to the damping and $\Gamma$ is always positive (see both Sec.~\ref{sec:nl_freq_damp}).
For the model we consider here, $\Gamma$ has a particularly small value of $\Gamma/\omega_{\rm prnt}\sim 10^{-9}$. For Jupiters with $R > 1.1 R_{\rm J}$, $\Gamma$ may be significantly larger. 
In part, this is because the damping rate due to turbulent convection increases sharply with increasing  $R$~\citep{Arras:09}. In addition, we find that irradiation causes a thin radiative zone to form near the surface where the daughter p-modes' shears peak. This further reduces the dissipation due to turbulent convection compared to the case without irridiation.

From this point onward, we will use values shown in Table~\ref{tab:models} as the default parameters for the planetary model. A primary goal of this paper is to develop the theoretical framework for diffusive tidal evolution including nonlinear mode interactions. A more comprehensive survey on how the tidal evolution trajectories depends on different values of $(\Omega, \Gamma)$, and how $(\Omega, \Gamma)$ further vary with respect to $(M, R)$ is deferred to future work. 

\section{Iterative map including nonlinear effects}
\label{sec:iter_map}

We now have the ingredients to perform an iterative map similar to the one used by~\citet{Vick:18} but now including nonlinear effects.
Since we consider here a single parent mode $a$, and the daughters' effects are collectively absorbed into $\Omega$ and $\Gamma$, we will drop the subscript $a$ in mode amplitude and energy from this point onward.%\nevin{Or just in this section?} 

Suppose the parent mode has an amplitude  $q_{k-1}'^{(1)}$ in the inertial frame right before the $k$'th pericenter passage. Its amplitude right after the $k$'th passage, $q_k'^{(0)}$, is given by Eq.~(\ref{eq:iter_map_interation}), just as in the linear case (but see the discussion below of the potential impact of nonlinear effects on the parent's kick).
Given $q_k'^{(0)}$,  the orbital energy and period of the $k$'th cycle are given by Eqs.~(\ref{eq:iter_map_E_orb}) and~(\ref{eq:iter_map_P_orb}), respectively. 

The next step of the mapping is to relate $q_k'^{(0)}$ to $q_{k}'^{(1)}$, the amplitude right before the $(k+1)$'th passage.  While the model we consider in Sec.~\ref{sec:values_for_real_jupiters} has a particularly weak dissipation, to obtain the accumulated phase over the course of the $k$'th orbit one needs to account for the gradual decay of the parent's energy due to linear and nonlinear damping. 
This can be achieved by first obtaining the energy $\tilde{E}_{k}^{(1)} = |q_{k}'^{(1)}|^2$ using the second line of Eq.~(\ref{eq:E_vs_t}) with $\tilde{E}_a^{(0)}=\tilde{E}_k^{(0)}$ and $t=P_{{\rm orb}, k}$. The evolution phase right before the $(k+1)$'th passage is then 
\begin{equation}
    \phi_k = -\omega_a' P_{{\rm orb}, k} + \delta \phi_{{\rm nl},k}.
    \label{eq:phi_k}
\end{equation}
Here $\delta \phi_{{\rm nl},k}$ is the excess phase due to the nonlinear frequency shift (the frequency shift due to linear damping is negligible since $\gamma_a \ll \omega_a'$), which can be calculated using Eq.~(\ref{eq:dphi_vs_t_leading_order_small_diss}) with $\tilde{E}_a^{(0)}=\tilde{E}_k^{(0)}$ and $\tilde{E}_a =\tilde{E}_k^{(1)}$.\footnote{If a system is strictly conservative, then we have $\tilde{E}_k^{(1)}=\tilde{E}_k^{(0)}$, and one can use  Eq.~(\ref{eq:dphi_vs_t_leading_order_small_diss}) with $t=P_{\rm orb, k}$ to obtain the phase. }

Before we proceed, it is important to point out a few caveats to this approach. 
First of all, the expressions we derive in this work are only the leading-order nonlinear corrections. They are accurate only when the parent mode's energy satisfies $\tilde{E}_k \lesssim 10^{-3}$. Therefore, in this work our focus will be on \emph{the initial triggering of the diffusive tide by the nonlinear mode coupling, particularly the nonlinear phase shift $\propto \Omega$.} The evolution timescale we consider here is thus typically a few hundreds of years or less, when the parent mode is still building up its energy. We defer the examination of tidal evolution over $\sim 10\,{\rm kyr}$ to follow-up studies in this series, as such a study would require both modifications to our leading-order expressions, and energy dissipation mechanisms due to both weakly nonlinear damping $\propto \Gamma$ and strongly nonlinear wave-breaking as considered in \citet{Wu:18}.\footnote{We estimate that one would need $\Gamma/\omega_a\gtrsim 10^{-6}$ to prevent the parent mode from evolving into the wave-breaking regime $\tilde{E}_k\gtrsim 0.1$ by the weakly-nonlinear damping as we consider here. While this is much greater than the nonlinear damping rate we find for the $R=1.1\,R_{\rm J}$ model, it can nonetheless be achieved if the Jupiter has a greater radius. E.g., we find a Jupiter model $R=2.0\,R_{\rm J}$ can have $\Gamma/\omega_a\simeq 5\times10^{-4}$. }  

Secondly, we assumed that the daughters' amplitudes are given by their instantaneous steady-state values, Eqs.~(\ref{eq:q_b_ss_soln}) and (\ref{eq:q_c_ss_soln}). 
We show in Appendix~\ref{sec:q_b} that this may not be strictly true if a daughter mode $b$ (with $|m_b|=4$) has $|2\omega_a + \omega_b|\lesssim \Omega_{\rm peri}$. Specifically, there should be an additional contribution to the daughter's amplitude that depends on the past history of the mode network. Nonetheless, we drop such corrections for simplicity in the current study. Our numerical experiments suggest this term becomes potentially important only after a few thousand  orbital cycles, and therefore should not affect the initial triggering of the diffusive process we consider in this work. 

Lastly, we have assumed that the ``kick'' at each pericenter passage is always given by the linear calculation. In reality, the kick $\Delta q_1$ depends on the parent's eigenfrequency [through $K_{lm}$; Eq.~(\ref{eq:K_lm_para})] which changes nonlinearly. Consequently, $\Delta q_1$ should also be modified by the nonlinear frequency shift. However, for $\tilde{E}_k\lesssim 10^{-3}$, the fractional decrease of the parent's eigenfrequency $\Omega E\tilde{E}_k/\omega_a$ is only a few percent.\footnote{By comparison, the fractional change in $\Omega_{\rm peri}$ is only $\mathcal{O}(10^{-5})$ as the mode energy grows from 0 to $\tilde{E}_k\lesssim 10^{-3}$; see Sec.~\ref{sec:ang_momentum}. }
The change in the one-kick amplitude is thus less than $20\%$ according to Eq.~(\ref{eq:K_lm_para}). Its effect can be more significant as the parent mode's energy further builds up, however; we plan to examine this in follow-up studies.

\section{Triggering diffusive growth}
\label{sec:trigger_growth}

\subsection{Relative importance of linear and nonlinear effects for triggering diffusive growth}
\label{sec:relative_importance}
The main question we want to investigate in this paper is how do nonlinear mode interactions affect the threshold for triggering the  diffusive growth of the f-mode?  In order to trigger diffusive growth, the phase evolution of the mode must vary randomly from orbit to orbit by an amount~\citep{Vick:18, Wu:18}
\begin{equation}
    |\Delta \phi_{k}| = |\phi_k - \phi_{k-1}| > \mathcal{O}(1)\,{\rm rad},
    \label{eq:dphi_k_def}
\end{equation}
where $\phi_k$ is given by Eq.~(\ref{eq:phi_k}). 
In linear theory, this is achieved through the tidal back-reaction on the orbit. At each passage, a random amount of energy $\Delta \tilde{E}_k$ is removed from the orbit, which changes the orbital period by $\Delta P_{{\rm orb}, k}$ [Eq.~(\ref{eq:iter_map_P_orb})]
and consequently the phase by $\Delta \phi_{{\rm br}, k}=-\omega_a'\Delta P_{{\rm orb}, k}$, where the subscript ``br'' stands for ``back-reaction''. However, linear theory neglects the fact that the energy $\Delta \tilde{E}_k$ gained by the planet's f-mode also changes its eigenfrequency by $\Delta \omega_a(\Delta \tilde{E}_k)=\delta \omega_a(\tilde{E}_k) - \delta \omega_a(\tilde{E}_{k-1})\simeq \Omega \Delta \tilde{E}_{k}$. This frequency shift induces an additional random phase variation  (relative to the previous orbit) $\Delta \phi_{{\rm nl}, k}\simeq -\left(\Omega \Delta \tilde{E}_{k}\right) P_{{\rm orb}, k}$. 
The nonlinear frequency shift therefore provides another way of triggering the f-mode's diffusive growth.

Quantitatively, \citet{Vick:18} found that it is sufficient to consider the phase shift after the first pericenter passage in order to determine the boundary for diffusion to happen. Specifically, let $\Delta \tilde{E}_1\equiv |\Delta q_1|^2$ be the energy gained by the mode after the first pericenter passage (suppose it starts with an amplitude $|q_0|\ll |\Delta q_1|$). The phase shift caused by the tidal back-reaction after the first pericenter passage is thus\footnote{To be consistent with the indexing convention used in Eq.~(\ref{eq:iter_map_P_orb}), we should use $P_{\rm orb, 1}$ and $\tilde{E}_{\rm orb, 1}$ in Eqs.~(\ref{eq:dphi_br_1}) and (\ref{eq:dphi_nl_1}). Nonetheless, using the quantities evaluated at cycle ``0'' will only cause a difference of $\mathcal{O}(\Delta \tilde{E}_1^2)\ll 1$, which can be safely ignored. } 
\begin{equation}
    \Delta \phi_{\rm br, 1} = -\omega_a' \Delta P_{\rm orb, 1} =\frac{3}{2}\omega_a' P_{\rm orb, 0}\frac{\Delta \tilde{E}_1}{\big{|}\tilde{E}_{\rm orb, 0}\big{|}},
    \label{eq:dphi_br_1}
\end{equation}
where  
in the second equality we have used the fact that  $\tilde{E}_{\rm orb, 0}=E_{\rm orb,0}/E_0<0$. The threshold for growth is approximately $|\Delta \phi_{\rm br, 1}|\simeq 1\,{\rm rad}$~\citep{Vick:18}, 
which corresponds to a threshold one-kick energy $\Delta \tilde{E}_1$ of 
\begin{eqnarray}
    \Delta \tilde{E}_{\rm br, 1} &\simeq& 1.0\times10^{-5}
    \left(\frac{\omega_a'}{\omega_a}\right)^{-1}
    \left(\frac{\omega_a}{1.1\omega_0}\right)^{-1}
    \left(\frac{R}{1.1R_{\rm J}}\right)^{5/2}
     \nonumber \\
    &&\times \left(\frac{M}{M_{\rm J}}\right)^{-3/2}\left(\frac{M_\ast}{M_\odot}\right)^{3/2}\left(\frac{a_{\rm orb,0}}{{\rm AU}}\right)^{-5/2}. 
    \label{eq:dE1_thres_br}
\end{eqnarray}

The nonlinear frequency shift also leads to an excess phase of 
\begin{equation}
    \Delta \phi_{\rm nl, 1}\simeq - \left(\Omega\Delta \tilde{E}_1\right) P_{\rm orb, 0} ,
    \label{eq:dphi_nl_1}
\end{equation}
where the subscript ``nl'' stands for ``nonlinear'' effects, and we have used Eq.~(\ref{eq:domega_leading_order}) for the nonlinear frequency shift. 
By setting $|\Delta \phi_{\rm nl, 1}|=1\,{\rm rad}$ we similarly obtain the one-kick energy threshold to trigger diffusive growth soley from nonlinear effects,
\begin{eqnarray}
    \Delta \tilde{E}_{\rm nl, 1} &\simeq& 1.8\times 10^{-6}\left(\frac{|\Omega|}{30\omega_a}\right)^{-1}\left(\frac{\omega_a}{1.1\omega_0}\right)^{-1} \left(\frac{R}{1.1R_{\rm J}}\right)^{3/2}
    \nonumber  \\
    &&\times \left(\frac{M}{M_{\rm J}}\right)^{-1/2}\left(\frac{M_\ast}{M_\odot}\right)^{1/2}
    \left(\frac{a_{\rm orb,0}}{{\rm AU}}\right)^{-3/2},
    \label{eq:dE1_thres_nl}
\end{eqnarray}
where we plugged in values representative of the hot Jupiter model described in Section~\ref{sec:values_for_real_jupiters}.  Comparing Equations~(\ref{eq:dE1_thres_br}) and (\ref{eq:dE1_thres_nl}),  we see that the nonlinear frequency shift can have a significantly lower one-kick  energy threshold than that of tidal back-reaction.  It can therefore play a critical role in triggering the diffusive growth of a mode. Furthermore, as we show in Sec.~\ref{sec:values_for_real_jupiters} that for realistic Jupiter models,  $\Omega<0$ typically and thus $\Delta \phi_{\rm br, 1} \Delta \phi_{\rm nl, 1}>0$.  
Intuitively, this can be understood as  follows. Suppose a positive amount of energy is transferred from the orbit to the mode, and as a result the orbital period decreases. Meanwhile, this energy lowers the frequency at which the mode oscillates as $\Omega < 0$ for typical Jupiter models (Table~\ref{tab:models}). Consequently, both effects make $\omega_a P_{\rm orb}$ decrease.
We thus see the two effects add together to further lower the threshold. 

In order to better see how the relative importance of the two effects scales with the various parameters, we take their ratio
\begin{eqnarray}
    \frac{|\Delta \phi_{\rm nl, 1}|}{|\Delta \phi_{\rm br, 1}|} &=& \frac{2}{3}\frac{|\Omega|}{\omega_a} \frac{\omega_a}{\omega_a'} \frac{|E_{\rm orb,0}|}{E_0}, \nonumber \\
    &=&5.6
    \left(\frac{|\Omega|}{30\omega_a}\right)
    \left(\frac{\omega_a'}{\omega_a}\right)^{-1}
    \left(\frac{R}{1.1R_{\rm J}}\right) \nonumber \\
    &&\times \left(\frac{M}{M_{\rm J}}\right)^{-1}\left(\frac{M_\ast}{M_\odot}\right)
    \left(\frac{a_{\rm orb,0}}{{\rm AU}}\right)^{-1}. 
    \label{eq:phi_ratio}
\end{eqnarray}
Note that the ratio is  independent of $\Delta \tilde{E}_1$ and $P_{\rm orb, 0}$.  Instead, it mainly depends on the ratio of orbital energy to binding energy of the planet, $\sim (M_\ast/M)(R/a_{\rm orb})$.
Consequently, as $a_{\rm orb}$ decreases during the orbital circularization process, 
the nonlinear phase shift becomes increasingly dominant over the back reaction shift.
This suggests that the  nonlinear frequency shift will play a crucial role in maintaining the diffusive energy transfer from the orbit to the planetary mode in the circularization process.

\subsubsection{Significance of nonlinear effects in other types of eccentric binaries}

We can use Eq.~(\ref{eq:phi_ratio}) to also estimate the significance of the nonlinear effects in other binary systems with highly eccentric orbits. For a binary of solar-type stars in a highly eccentric orbit with $a_{\rm orb}\sim {\rm AU}$, the nonlinear phase shift is only $~\sim 1\%$ of that due to tidal back-reaction.\footnote{
This is specifically for the nonresonant nonlinear effect of the $l_a=2$ f-mode. A solar-type star also has low-frequency g-modes that have very different values of $\Omega$ from that of the f-mode. Those g-modes may  allow for a richer family of nonlinear effects, such as the parametric instability.}  
A similar ratio of a few percent is also found for a neutron star binary  with $a_{\rm orb} \simeq 1000\,{\rm km}$~\citep{Vick:19b}. Indeed, both solar-type stars and neutron stars are more compact (i.e., with smaller $M/R$) than a typical Jovian planet.
Therefore, an eccentric hot Jupiter offers an especially interesting  system for  studying the impact of  nonlinear effects on diffusive growth.

\begin{figure}
   \centering
   \includegraphics[width=0.45\textwidth]{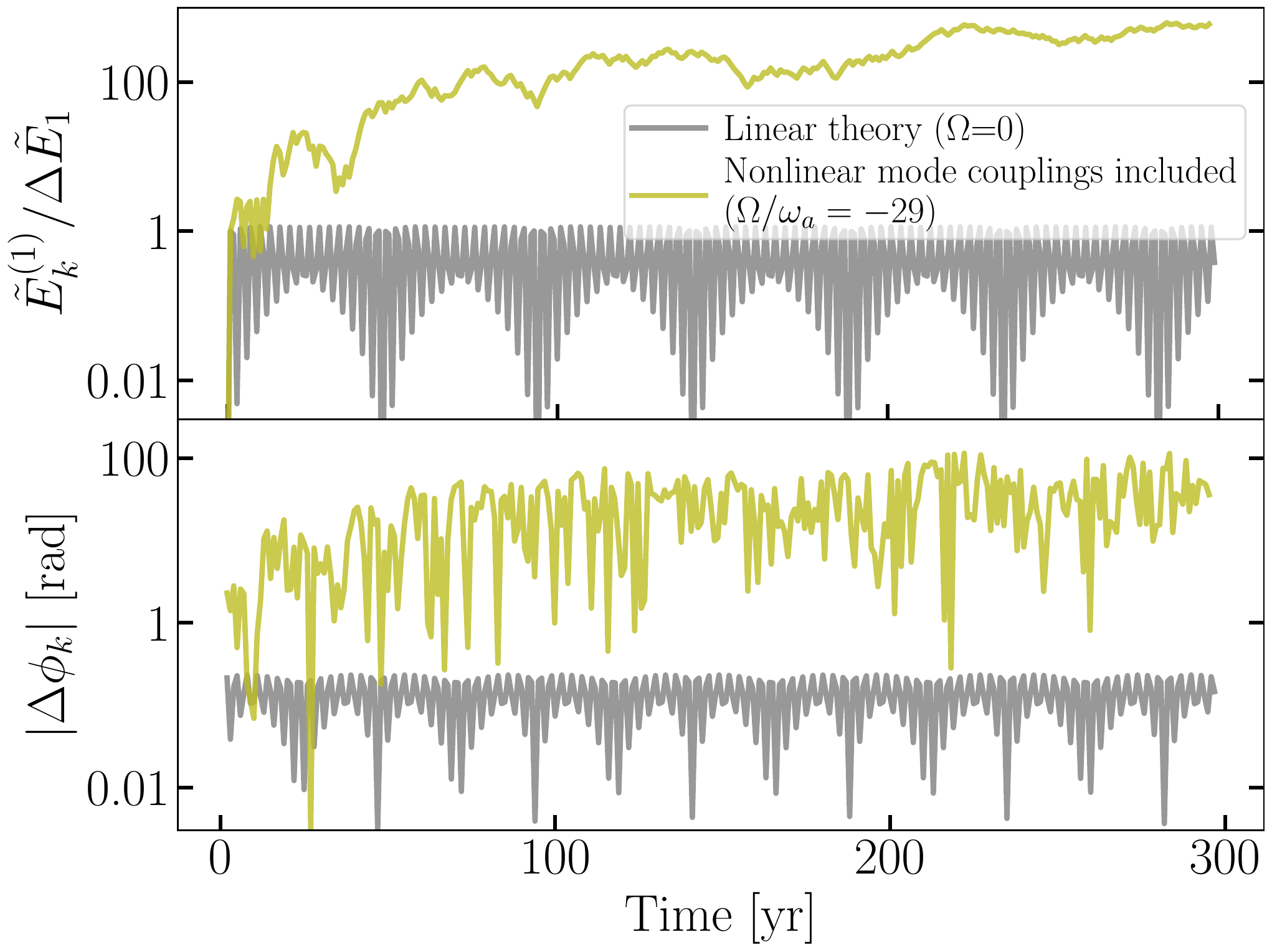} % requires the graphicx package
   \caption{
   Evolution trajectories during the first 300 orbits. The top panel shows the mode energy relative to the one kick energy and the bottom panel shows the difference in the mode's excess phase between the $k$'th and $(k-1)$'th orbit. We fix the pericenter distance at $D_{\rm peri}=3.95\,D_{\rm t}$ and assume the planet is non-rotating, resulting in a one-kick energy $\Delta \tilde{E}_1=4.5\times10^{-6}$. In the linear case ($\Omega=0$; grey lines) the mode energy and excess phase oscillate periodically and there is no diffusive growth.  However, when nonlinear mode interactions are included ($\Omega/\omega_a=-29$; olive lines)
   the excess phase is  significantly larger and varies randomly, resulting in diffusive growth of the f-mode's energy. 
    } 
   \label{fig:traj_LvsNL_comp_lo}
\end{figure}

\subsection{Early orbital evolution following the onset of  diffusive growth}
In Fig.~\ref{fig:traj_LvsNL_comp_lo} we show a  representative example of the first 300 orbits of a Jovian planet orbiting a solar-type star at a semi-major axis $a_{\rm orb}=1\,{\rm AU}$ using the iterative map described in Sec.~\ref{sec:iter_map}.  The planet's  parameters are given in Table~\ref{tab:models} and we assume here that the planet is not rotating.  In the figure, the pericenter distance is set to $D_{\rm peri}=3.95\,D_{\rm t}$, where $D_{\rm t}\equiv R (M_\ast/M)^{1/3}\simeq 11 R_{\rm J}\simeq 5.3\times10^{-3}\,{\rm AU}$ is the tidal radius of the planet.
The corresponding eccentricity is thus $e_{\rm orb} = 1 - D_{\rm peri} / a_{\rm orb} = 0.979$.
For these parameters, the one-kick energy is $\Delta \tilde{E}_1=|\Delta q_1|^2=4.5\times10^{-6}$ [Eq.~(\ref{eq:one_kick_amp_vs_Klm})]. 
The top panel  shows the energy of the parent mode (with $l_a=m_a=2$) and the bottom panel shows the difference of the evolution phase between two adjacent cycles [Eq.~(\ref{eq:dphi_k_def})]. We see that in the linear case (grey trace), the  difference in the mode's propagation phase between adjacent cycles, $|\Delta \phi_{\rm k}|$, is small ($\sim 0.1\,{\rm rad}$) and the mode energy just undergoes periodic oscillations (see also the discussions in \citealt{Vick:18}). By contrast, when we include nonlinear mode interactions there is an additional contribution to the random phase due to the nonlinear frequency shift [Eq.~(\ref{eq:dphi_nl_1})].  As a result, we see that the f-mode's energy grows diffusively, unlike in the linear case.   After 300 cycles, the mode energy grows to about $300\Delta \tilde{E}_1$, as one would expect for a diffusive process (i.e., the amplitude grows as the square root of the number of pericenter kicks). 

\begin{figure*}
\subfloat[Non-spinning.]{
	\begin{minipage}[c][0.98\width]{0.47\textwidth}
	   \centering
	   \includegraphics[width=0.95\textwidth]{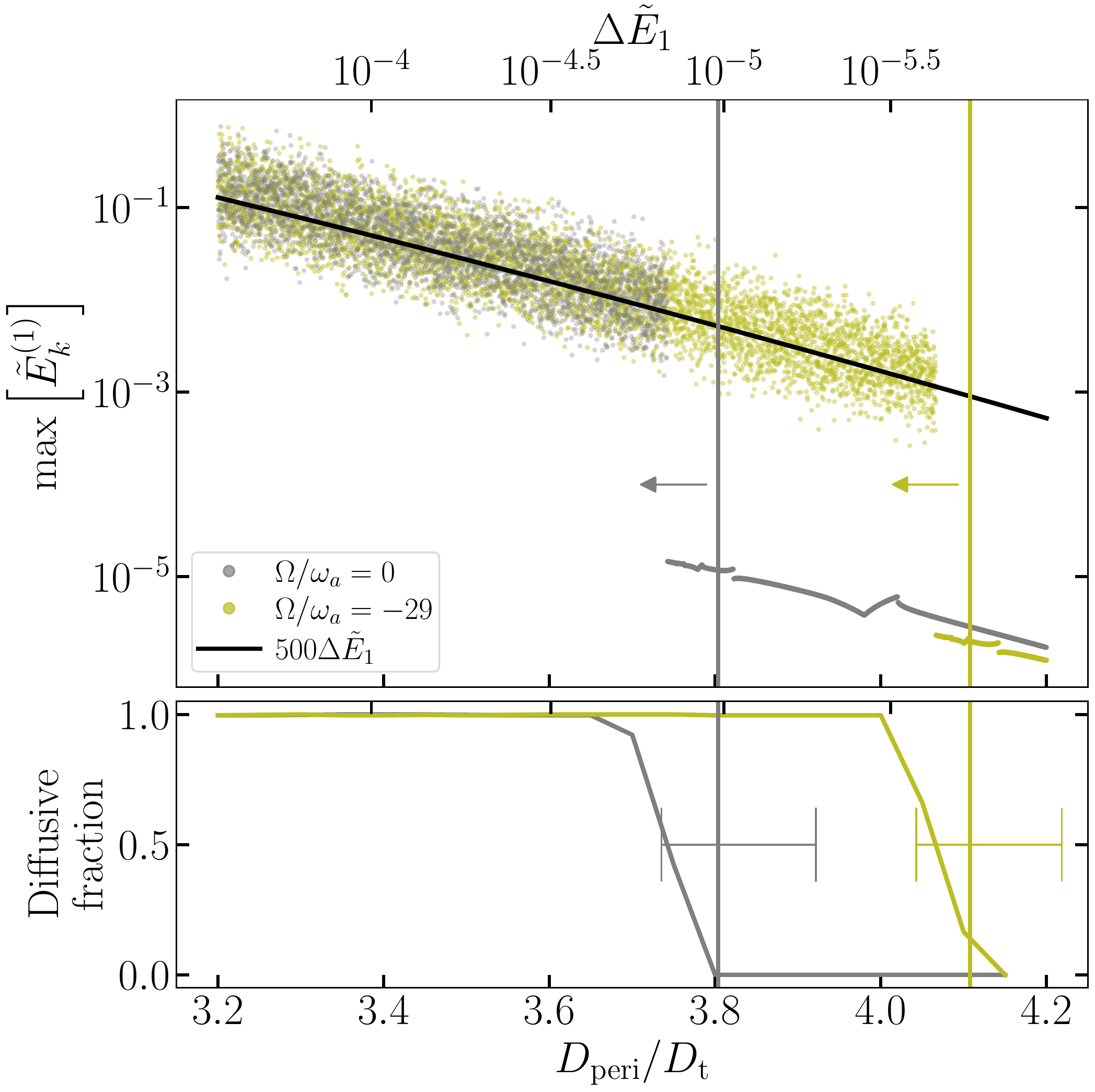}
	\end{minipage}}
 \hfill 	
  \subfloat[Pseudo-synchronized.]{
	\begin{minipage}[c][0.98\width]{0.47\textwidth}
	   \centering
	   \includegraphics[width=0.95\textwidth]{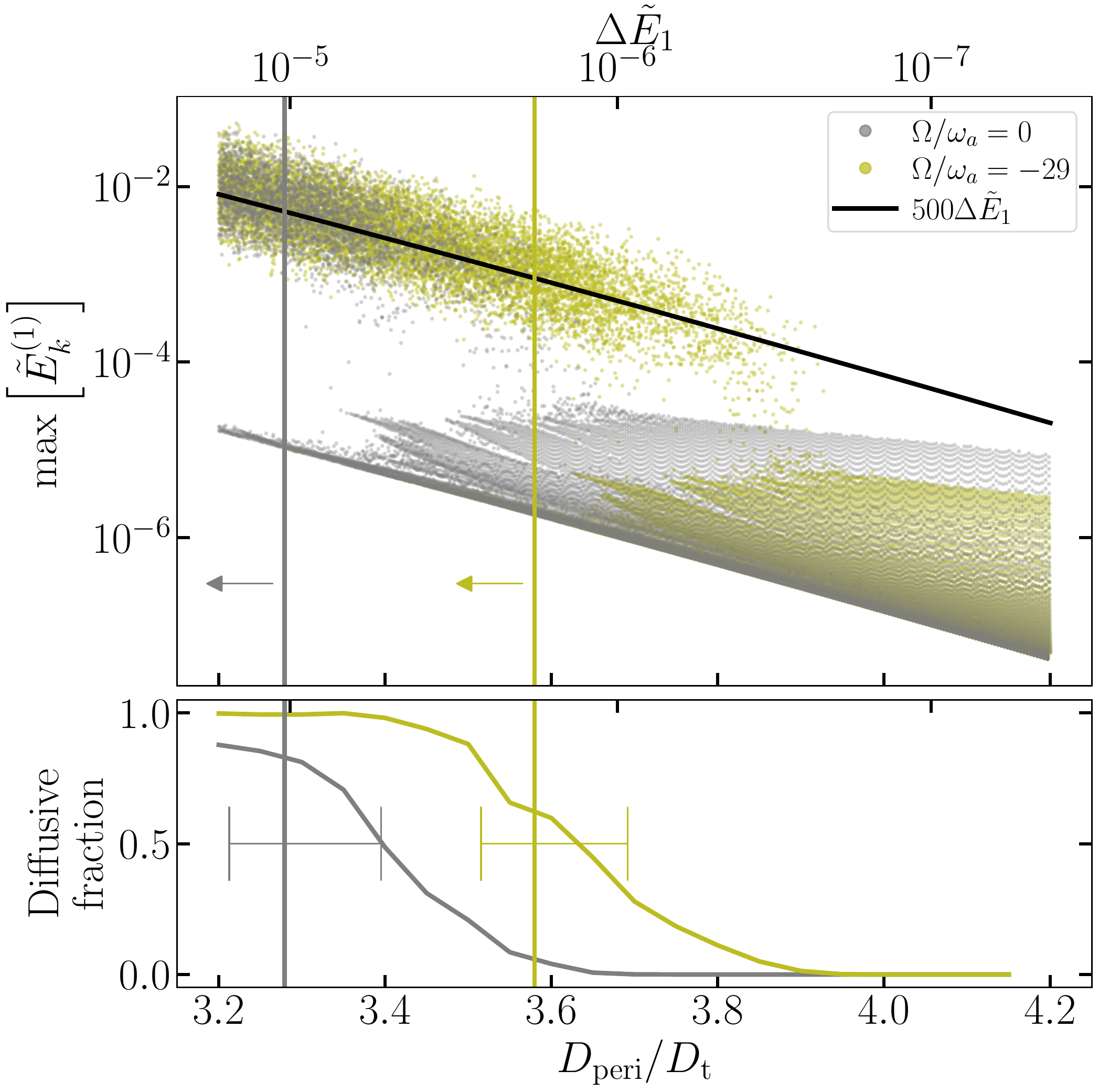}
	\end{minipage}}
\caption{Top panels: maximum f-mode energy achieved after 500 pericenter passages. The grey circles are calculated assuming linear theory while the olive circles also include nonlinear mode interactions (with $\Omega/\omega_a=-29$). In both plots we set $a_{\rm orb,0}=1\,{\rm AU}$. In the left panel we assume that the planet is not spinning while in the right panel we assume that it spins at a constant rate given by the pseudo-synchronization condition [Eq.~(\ref{eq:omega_s_ps})]. 
Bottom panels: the fraction of systems that undergoes the diffusive growth (i.e., points around the black lines) as a function of $D_{\rm peri}$, estimated over a full bin width of $0.1\, D_{\rm t}$. 
The vertical lines are the analytic estimates for the diffusive growth threshold based on Eqs.~(\ref{eq:dE1_thres_br}) and (\ref{eq:dE1_thres_nl}) by setting $\Delta \phi_{1}=1\,{\rm rad}$. The error bars are obtained if we instead use  $\Delta \phi_{1}=0.5\,{\rm rad}$ or $1.5\,{\rm rad}$.
The region where we expect diffusive growth to occur are also indicated by arrows in the top panel. 
In the right panel, a mode to the right of the vertical lines can occasionally grow to an intermediate amplitude of $\tilde{E}_{k}^{(1)}\simeq \text{a few}\times 10^{-5}$ if its frequency $\omega_a'$ comes to close resonance with one of the orbital harmonics. 
\label{fig:maxE_vs_Dperi}}
\end{figure*}

In Fig.~\ref{fig:maxE_vs_Dperi} we systematically explore some of the conditions necessary to trigger  diffusive growth. We show the maximum mode energy achieved after 500 orbital cycles (about 500 years) as a function of the pericenter distance [or equivalently, the orbital eccentricity since $D_{\rm peri}=a_{\rm orb}(1-e_{\rm orb})$ and we set the initial semi-major axis at $a_{\rm orb,0}=1\,{\rm AU}$]. For the given planetary model (Table~\ref{tab:models}), the one-kick energy $\Delta \tilde{E}_1$ is shown in the top x-axis for each choice of $D_{\rm peri}$. In the left panel, we assume the planet is non-rotating, while in the right panel we assume it is pseudo-synchronized with the orbit with $\Omega_{\rm s} / \omega_a  = 0.19 (D_{\rm peri}/0.02\,{\rm AU})^{-3/2}$ and $\omega_a'=\omega_a + 2\Omega_{\rm s}$. If a mode experiences diffusive growth, then its energy after $k$ pericenter passages is expected to be $\tilde{E}_{\rm k}\sim k \Delta \tilde{E}_{1}$ on average. Since we set $k=500$, we expect $\max\left[\tilde{E}_k\right]\simeq 500 \Delta \tilde{E}_1$ for a mode that grows diffusively (indicated by the black lines), while $\max\left[\tilde{E}_k\right]\sim \Delta \tilde{E}_1$ for a mode that does not grow (assuming the mode is off resonance with the orbit; in the right panel $\omega_a'$ changes as we vary $D_{\rm peri}$, allowing it to scan through a series of resonances with different orbital harmonics, thereby causing the excess features to the right of the vertical lines, which we will discuss in Sec.~\ref{sec:diffusive_inc_spin}). 

We see that each panel in Fig.~\ref{fig:maxE_vs_Dperi} can be divided up into two regions according to the maximum mode energy achieved. Let us first focus on the left panel (a non-rotating planet 
with $\omega_a P_{\rm orb,0}/2\pi=2818.73$ being a non-integer
). In the linear case ($\Omega=0$), we find 
numerically
that the boundary where diffusive growth is first triggered is at  
$D_{\rm peri}\simeq 3.75D_{\rm t}$ , corresponding to a one-kick energy of $\Delta \tilde{E}_1\simeq 1.4\times 10^{-5}$.  
The analytical estimate [Eq.~(\ref{eq:dE1_thres_br}); see vertical grey line], agrees well with the numerical results but slightly overestimates the threshold value of $D_{\rm peri}$ (and underestimates $\Delta \tilde{E}_{1}$) because there we simply assumed the threshold phase shift to be 1\,rad; in reality a slightly greater phase shift is required.
This can also be seen from the bottom panel where we show the fraction of systems undergoing diffusive growth (i.e., the fraction of points around the black lines; the estimate is preformed over a full bin width of $0.1 D_{\rm t}$). The error bars around the vertical lines are obtained by setting $\Delta \phi_{1}=0.5\,{\rm rad}$ and $1.5\,{\rm rad}$ and then re-evaluating $\Delta \tilde{E}_1$ using Eqs.~(\ref{eq:dphi_br_1}) and (\ref{eq:dphi_nl_1}).

When we  account for the nonlinear frequency shift, we find that the boundary moves to larger $D_{\rm peri}$ (smaller one-kick energies). For the representative value of $\Omega/\omega_a\simeq -29$ (see Table~\ref{tab:models}), we find that the threshold one-kick energy is lowered to $\Delta \tilde{E}_1\simeq 2.4\times 10^{-6}$, which is a factor of about 6 smaller than  the linear case [see the vertical olive line and Eq.~(\ref{eq:dE1_thres_nl}); note that the threshold is in fact  determined by the sum $\Delta \phi_{\rm nl, 1}+\Delta\phi_{\rm br, 1}$, with the latter being $\simeq 20\%$ of the former for the parameters in Fig.~\ref{fig:maxE_vs_Dperi}]. Because the one-kick energy depends sensitively on the pericenter distance, a factor of six change in $\Delta \tilde{E}_1$ corresponds to a $\simeq 10\%$ increase in $D_{\rm peri}$. 

\subsection{Including spin effects}
\label{sec:diffusive_inc_spin}
We consider the effects of planet spin in the right panel of Fig.~\ref{fig:maxE_vs_Dperi}. We  assume the planet is rotating at a rate determined by the pseudo-synchronization condition [Eq.~(\ref{eq:omega_s_ps})].\footnote{
We assume pseudo-synchronization here as a plausible scenario. Whether it can be achieved through, e.g., a Lidov-Kozai evolution involving a tertiary mass, remains to be answered by future studies. 
}  
As we show in Sec.~\ref{sec:lin_prob}, the mapping equations including spin are formally the same as the non-spinning equations except that the mode frequency is replaced by the inertial frame value $\omega_a' = \omega_a + m_a \Omega_{\rm s}$ (this frequency then enters the calculations of the one-kick amplitude and the linear propagation phase). 
Since we focus on a prograde mode with $m_a=2$, $|\Delta q_1|\propto K_{22}$ decreases sharply as $\omega_a'$ increases [Eq.~(\ref{eq:K_lm_para})].  As a result, the $D_{\rm peri}$ boundary where diffusive growth is first triggered is shifted to smaller values.

At the same time, in the right panel the mode can occasionally become resonant with the orbit when $\omega_a'P_{\rm orb}/2\pi={\rm integer}$, as $\omega_a'$ now varies with $D_{\rm peri}$ due to pseudo-synchronization condition
(this is in contrast to the left panel where $\omega_a P_{\rm orb}/2\pi$ is fixed at a non-integer value when we vary $D_{\rm peri}$). For $D_{\rm peri}/ D_{\rm t}\gtrsim 3.8$, such resonances can bring the mode energy up to $\tilde{E}^{(1)}_k\simeq \text{a few}\times10^{-5}\gg \Delta \tilde{E}_1$. However, as the mode acquires energy from the orbit, the orbital period starts to change (though not by a significant enough amount to trigger diffusion). It thus  destroys the resonance and prevents the mode energy from increasing further (see also \citealt{Vick:18}). Similarly, the nonlinear frequency shift also destroys the resonance between $\omega_a'$ and $P_{\rm orb}$ and this is why the upper envelope of the olive dots is at a lower value than that of the grey ones.

At $3.2\lesssim D_{\rm peri}/ D_{\rm t}\lesssim 3.8$, we see that the chance resonance with the orbit may occasionally help a slightly sub-threshold mode to also evolve into the diffusive regime. Suppose the chance resonance helps the mode to initially build up an energy $\tilde{E}_{k,0}\simeq 10^{-5} (10^{-4})$ with (without) the nonlinear effect (corresponding approximately to the upper envelopes shown in the right panel).  The typical energy exchange between a mode and the orbit is then given by $\sqrt{\tilde{E}_{k,0}\Delta \tilde{E}_1}\gg\Delta \tilde{E}_1$~\citep{Wu:18}. If we replace $\Delta \tilde{E}_1$ by $\sqrt{\tilde{E}_{k,0}\Delta \tilde{E}_1}$ in Eqs.~(\ref{eq:dphi_br_1}) and (\ref{eq:dphi_nl_1}), we see the new threshold one-kick energy becomes $\Delta \tilde{E}_{br,1}\simeq 10^{-6}$ and $\Delta \tilde{E}_{nl,1}\simeq 3\times10^{-7}$ for modes that are initially in close resonance with the orbit. As $D_{\rm peri}$ decreases and $\Delta \tilde{E}_1$ increases, even systems that are not at the upper envelope may enter the chaotic regime and the fraction of diffusive systems increases as indicated by the lower panel. Eventually, when $\Delta \tilde{E}_1$ reaches the value derived in Eqs.~(\ref{eq:dE1_thres_br}) and (\ref{eq:dE1_thres_nl}), almost all of the systems will grow diffusively.

\begin{figure*}
\subfloat[Non-spinning.]{
	\begin{minipage}[c][0.98\width]{0.47\textwidth}
	   \centering
	   \includegraphics[width=0.95\textwidth]{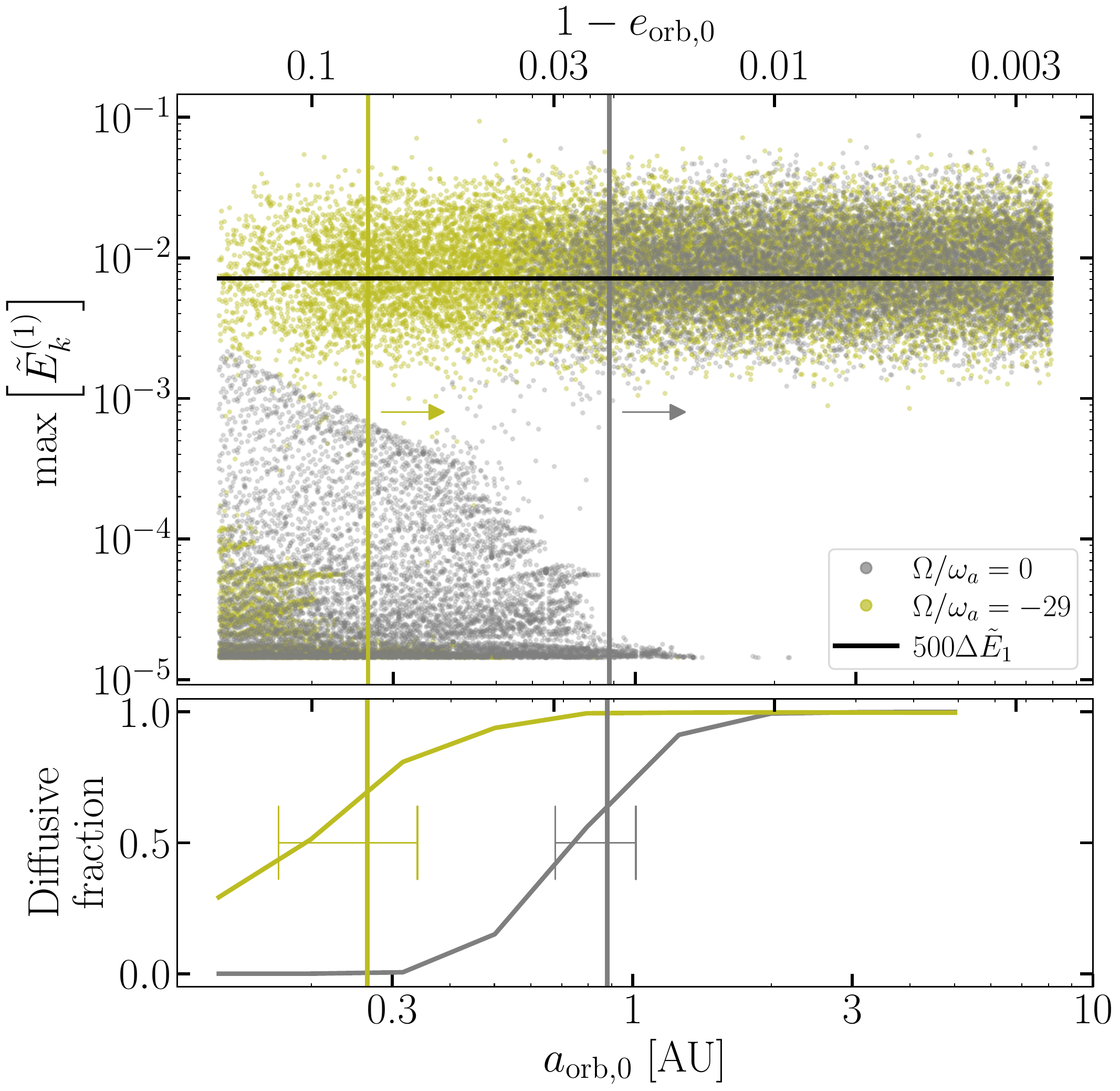}
	\end{minipage}}
 \hfill 	
  \subfloat[Pseudo-synchronized.]{
	\begin{minipage}[c][0.98\width]{0.47\textwidth}
	   \centering
	   \includegraphics[width=0.95\textwidth]{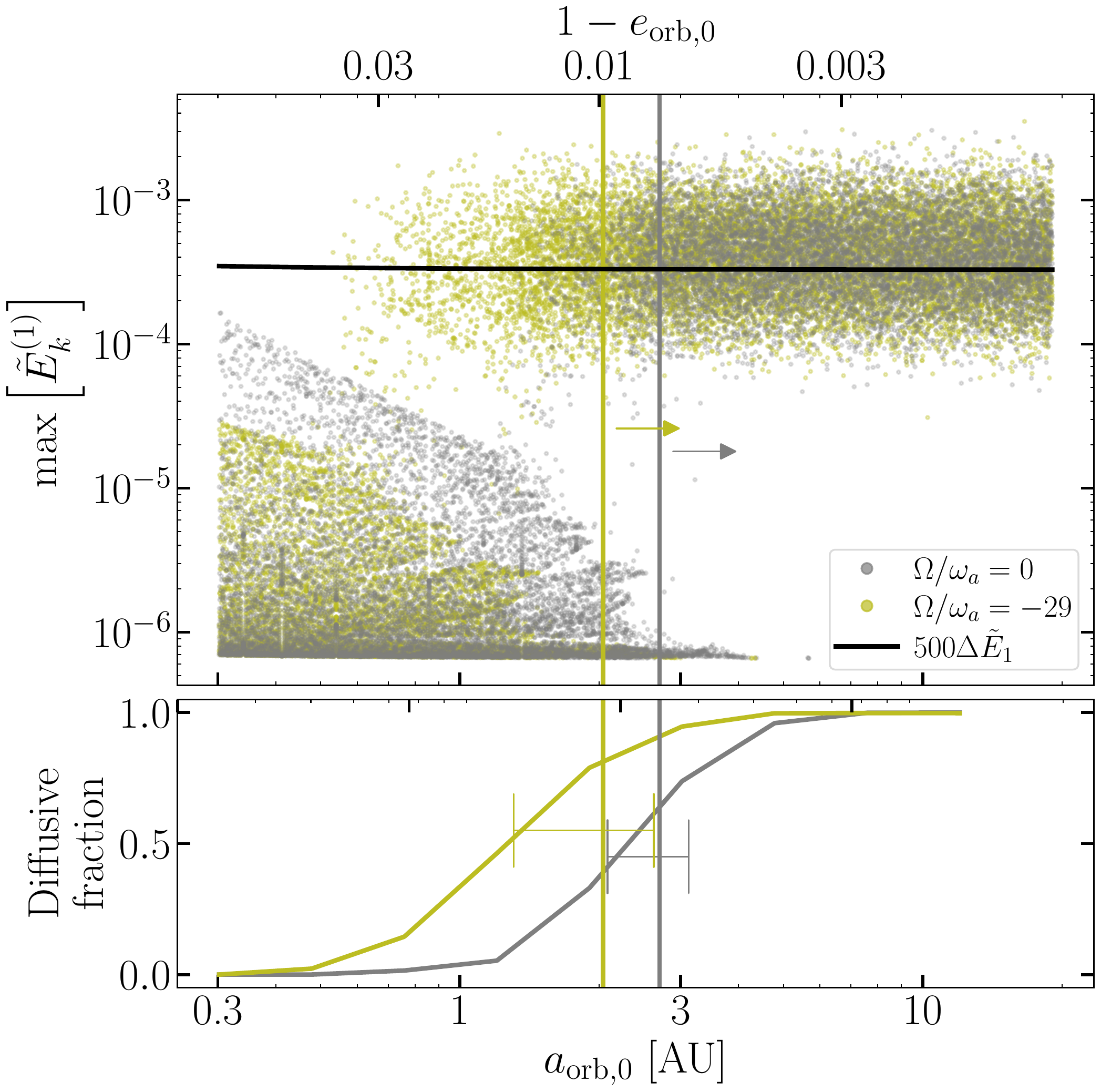}
	\end{minipage}}
\caption{Similar to Fig.~\ref{fig:maxE_vs_Dperi} but this time we fix the pericenter distance at $D_{\rm peri}=0.02\,{\rm AU}$ and let the initial semi-major axis $a_{\rm orb,0}$ vary (the top axis of each panel shows the corresponding $1-e_{\rm orb, 0}$). The grey and olive vertical lines are calculated using Eqs.~(\ref{eq:a0_thres_br}) and (\ref{eq:a0_thres_nl}),  respectively, assuming $\Delta \phi_1 = 1\,{\rm rad}$ is needed to trigger diffusive growth; the error bars show the threshold if instead $\Delta \phi_1 = 0.5\,{\rm rad}$ or $1.5\,{\rm rad}$ are needed. 
The fraction of systems undergoing diffusive growth shown in the bottom panel is estimated over logarithmic bins with full width of $\log_{10} \left(a_{\rm orb,0}/{\rm AU}\right)= 0.4$. 
A system with an $a_{\rm orb,0}$ that is slightly below the threshold may still trigger diffusive growth if the mode is close to resonance with the orbit. 
}
\label{fig:maxE_vs_a_orb}
\end{figure*}

\subsection{Threshold expressed in terms of semi-major axis rather than one-kick energy} 
An alternative way to consider the problem is to hold $D_{\rm peri}$ and thus $\Delta \tilde{E}_1$ fixed and instead vary the initial semi-major axis $a_{\rm orb, 0} = D_{\rm peri}/(1-e_{\rm orb, 0})$. By setting $|\Delta \phi_{\rm br, 1}|=1\textrm{ rad}$ in Eq.~(\ref{eq:dphi_br_1}) as before (Section~\ref{sec:relative_importance}) but now solving for $a_{\rm orb,0}^{\rm (br)}$, we find that the threshold to trigger diffusive growth due to only tidal backreaction is
\begin{eqnarray}
    a_{\rm orb,0}^{\rm (br)} &\simeq& 2.5\,{\rm AU} \left(\frac{\Delta \tilde{E}_1}{10^{-6}}\right)^{-2/5}
    \left(\frac{\omega_a'}{\omega_a}\right)^{-2/5}
    \left(\frac{\omega_a}{1.1\omega_0}\right)^{-2/5}
     \nonumber \\
    &&\times \left(\frac{R}{1.1R_{\rm J}}\right)\left(\frac{M}{M_{\rm J}}\right)^{-3/5}\left(\frac{M_\ast}{M_\odot}\right)^{3/5}.
    \label{eq:a0_thres_br}
\end{eqnarray}
Similarly, we can use Eq.~(\ref{eq:dphi_nl_1}) to find the threshold due to nonlinear mode interactions
\begin{eqnarray}
    a_{\rm orb,0}^{\rm (nl)} &\simeq& 1.5\,{\rm AU} \left(\frac{\Delta \tilde{E}_1}{10^{-6}}\right)^{-2/3}
    \left(\frac{|\Omega|}{30\omega_a}\right)^{-2/3}
    \left(\frac{\omega_a}{1.1\omega_0}\right)^{-2/3}
     \nonumber \\
    &&\times \left(\frac{R}{1.1R_{\rm J}}\right)\left(\frac{M}{M_{\rm J}}\right)^{-1/3}\left(\frac{M_\ast}{M_\odot}\right)^{1/3}.
    \label{eq:a0_thres_nl}
\end{eqnarray}
In both cases, the threshold $a_{\rm orb,0}$ \emph{increases} with decreasing one-kick energy $\Delta \tilde{E}_1$ (i.e., increasing $D_{\rm peri}$). 
This is because both $\Delta \phi_{\rm br}$ and $\Delta \phi_{\rm nl}\propto P_{\rm orb} \Delta \tilde{E}_1$ [Eq.~(\ref{eq:dphi_br_1}) and (\ref{eq:dphi_nl_1})], and a longer $P_{\rm orb} \propto a_{\rm orb} ^{3/2}$ is thus required in order for the f-mode to accumulate an excess phase $|\Delta \phi_k|$ to $\mathcal{O}(1)\,{\rm rad}$. Additionally, Eqs.~(\ref{eq:a0_thres_br}) and (\ref{eq:a0_thres_nl}) scale differently  with $\Delta \tilde{E}_1$ because $\Delta \phi_{\rm br}\propto |E_{\rm orb}|^{-1}$, which reflects the fact that an orbit with greater $a_{\rm orb}$ is less bound and thus sees a greater change in the fractional orbital period.  
Also note that Eq.~(\ref{eq:a0_thres_nl}) overestimates the minimum $a_{\rm orb,0}$ required to trigger the diffusion because it assumes only the nonlinear contribution to $\Delta \phi_{\rm k}$. In reality, the nonlinear frequency shift and the tidal back-reaction both contribute to $\Delta \phi_{\rm k}$ and they have the same sign (Section~\ref{sec:relative_importance}). 
Similar to the case shown in the right panel of Fig.~\ref{fig:maxE_vs_Dperi}, Eqs.~(\ref{eq:a0_thres_br}) and (\ref{eq:a0_thres_nl}) should be treated as the upper end of the thresholds; almost all of the systems with $a_{\rm orb, 0}$ greater than the values estimated in Eqs.~(\ref{eq:a0_thres_br}) and (\ref{eq:a0_thres_nl}) will trigger the diffusive evolution. On the other hand, if a system is initially close to being resonant with the orbit, then at smaller $a_{\rm orb, 0}$ it may still enter the diffusive regime. 

We evaluate the boundary in $a_{\rm orb, 0}$ numerically in Fig.~\ref{fig:maxE_vs_a_orb}. Here we fix the pericenter distance to be $D_{\rm peri} = 0.02\, {\rm AU}$, corresponding to a one-kick energy of $\Delta \tilde{E}_1\simeq 1.4\times10^{-5}$ ($\Delta \tilde{E}_1\simeq 6.7\times10^{-7}$) for a non-spinning (pseudo-synchronized) planet.
%\nevin{In the plots we say non-spinning and here non-rotating.  Obviously a very minor point but better to stick with one or the other.  Personally I like non-rotating better but either is  fine.} 
The situation is particularly interesting astrophysically in the case where the planet's spin is pseudo-synchronized. For a relatively weak one-kick energy of $\Delta \tilde{E}_1\lesssim 10^{-6}$, only planets born $\gtrsim 2\,{\rm AU}$ away from the host star can trigger diffusive tidal evolution and form hot Jupiters if only the linear theory is used. On the other hand, when we include nonlinear mode interactions, it can be triggered for planets born with $a_{\rm orb, 0}\simeq 0.7-2\,{\rm AU}$. 
Thus, nonlinear mode interactions significantly expand the parameter space allowed for diffusive growth to happen, which not only allows more potential progenitors to form hot Jupiters within the age of the Universe, but also saves more planets from disruption by the host star during the Kozai cycles (see, e.g., \citealt{Vick:19}). It could thus help alleviate the discrepancy between the predicted and observed hot Jupiter to regular Jupiter occurrence rate (see \citealt{Dawson:18}).

\section{Conclusion and Discussion}
\label{sec:conclusions}
We studied the nonlinear interaction between a self-coupled parent f-mode and daughter f- and p-modes in a Jovian planet. 
For a parent mode with azimuthal quantum number $m_a=2$ and frequency $\omega_a$, it drives both $m_b=-4$ daughters that correspond to waves oscillating at $2\omega_a$, and non-oscillatory $m_c=0$ daughters that correspond to a modification of the planet's structure (Sec.~\ref{sec:nonlinear}). 
We found that at leading order, the interaction leads to a nonlinear shift in the parent mode's eigenfrequency, $\delta \omega_a$, as well as a nonlinear increase in the parent mode's damping rate (imaginary part of the frequency), $\delta \gamma_a$. Both the nonlinear frequency shift and damping rate follow the scaling $\delta \omega_a, \delta \gamma_a\propto \omega_a E_a$ at leading order [Eqs.~(\ref{eq:domega_leading_order}) and (\ref{eq:dgamma_leading_order}); see also \citet{Kumar:94, Kumar:96}]. 
The modifications are time-dependent because we consider planets on highly eccentric orbits with parent mode energies $E_a$ that vary at each pericenter passage. 
Furthermore, we showed that although the frequency shift can, in principle, be either positive or negative, for typical Jupiter models a negative shift is more likely (that is, the parent mode's eigenfrequency decreases with increasing mode energy). The nonlinear damping, on the other hand, strictly increases as the mode energy increases (see Table~\ref{tab:models} and Fig.~\ref{fig:Omega_Gam_1p0Mj_1p1Rj}). 

We then developed the formalism to construct iterative maps including nonlinear effects and applied them to study how nonlinear interactions affect the high-eccentricity migration of proto-hot Jupiters. We found that the energy-dependent nonlinear frequency shift leads to an excess phase of the parent mode [Eq.~(\ref{eq:dphi_nl_1})] which is stochastic from orbit to orbit. It thus provides another channel for triggering the diffusive growth of the parent mode in addition to the tidal back-reaction considered in previous studies. 

In fact, we found that for typical Jupiter models, the nonlinear phase shift is $\approx 5$ times larger than the phase shift due to back-reaction [Eq.~(\ref{eq:phi_ratio})]. The two effects add together and lower the threshold one-kick energy required in order to trigger the growth by about a factor of $\approx 6$ compared to the case without nonlinear interactions (Fig.~\ref{fig:maxE_vs_Dperi}). Alternatively, if one fixes the one-kick energy, the threshold on the minimum initial orbital semi-major axis can be lowered by a factor of $\approx 2$ (Fig.~\ref{fig:maxE_vs_a_orb}). If the one-kick energy is small (due to either a small eccentricity  and hence large pericenter distance, or a high spin rate of the planet), then in the linear case only planets born at $a_{\rm orb,0} \gtrsim 2\,{\rm AU}$ can undergo diffusive tidal evolution and form hot Jupiters; however, when nonlinear interactions are accounted for, it is lowered to the interesting range of $a_{\rm orb,0} = 0.7-2\,{\rm AU}$.

In this paper, we focused on developing the theoretical framework and considered only the evolution over the first $\mathcal{O}(100)\,{\rm yr}$. There are several aspects of the problem we think would be interesting to address in future studies.

First, what is the long-term evolution of the system over $\sim 10\,{\rm kyr}$? For the Jupiter model we considered in this work (Table~\ref{tab:models}), the weakly nonlinear damping $\propto \Gamma$ is weak and thus the parent mode energy will grow so large that it likely becomes strongly nonlinear, as assumed by \citeauthor{Wu:18} (but see discussion below).  \citeauthor{Wu:18} found that the diffusive process, and hence the orbital evolution, typically stalls when the semi-major axis decays to $a_{\rm orb}\simeq 0.2\,\textrm{ AU}$ while the eccentricity is still high ($e_{\rm orb}\simeq 0.9$), and it was unclear what drives the subsequent orbital circularization. However, nonlinear mode interactions might prevent the circularization from stalling at high $e_{\rm orb}$ because the magnitude of the random phase it induces  decreases slower than that due to the tidal back-reaction as the orbit decays; see Eq.~(\ref{eq:phi_ratio}). 
This is because the orbit ``hardens'' ($|E_{\rm orb}|$ increases) as $a_{\rm orb}$ shrinks, which makes it increasingly hard to be perturbed by the tidal back-reaction [Eqs.~(\ref{eq:iter_map_P_orb}) and (\ref{eq:dphi_br_1})]. By contrast, the natural energy scale that enters the nonlinear phase shift is the $a_{\rm orb}$-independent binding energy of the planet (ignoring the evolution of the planet).
An efficient circularization could help explain both the paucity of super-eccentric Jupiters~\citep{Socrates:12} and the relatively young age of hot-Jupiter host stars~\citep{Hamer:19}. 

It would also be interesting to investigate  how the values of $(\Omega, \Gamma)$ vary for different Jupiter models and how the tidal evolution trajectories depend on $(\Omega, \Gamma)$. We estimate that if $\Gamma\gtrsim10^{-6}\omega_a$, weakly nonlinear damping could be sufficient to prevent the parent mode from evolving into the strongly nonlinear regime.  This would lead to another qualitative difference from the trajectories found in \citet{Wu:18}, in addition to the excess nonlinear phase shift discussed above. Such large dissipation rates can be achieved by Jupiter models with greater radii and it could have potentially important  observational consequences.

A calculation that combines diffusive tidal evolution with the mechanism that drives the eccentricity to large values in the first place (e.g., Lidov-Kozai cycles) 
would be valuable and help test these ideas further.  By including nonlinear effects, it would extend the work of   \citet{Vick:19} and thereby  provide a more robust estimate of the formation rate of hot Jupiters due to diffusive tidal evolution.  Current theories produce too few hot Jupiters relative to regular Jupiters, and it would be interesting to know whether nonlinear effects could help mitigate the tension.

To carry out the studies described above, a few modifications to the current framework would be needed. For instance, as the parent mode's energy builds up and its eigenfrequency decreases, the orbital integral $K_{lm}$ should be modified accordingly. Since the parent's frequency is typically shifted to a lower value ($\Omega<0$), we would expect the one-kick amplitude $\Delta q_{1}\propto K_{lm}$ to increase in magnitude as the parent's energy increases [Eq.~(\ref{eq:K_lm_para})]. This would further enhance the significance of the nonlinear effects. On the other hand, we do not expect a linear-in-energy frequency shift [Eq.~(\ref{eq:domega_leading_order})] to be accurate when $|\Omega|\tilde{E}_a\simeq \omega_a$. Note that this condition can happen at a smaller energy than the wave-breaking energy, and therefore further corrections would be needed. 

\acknowledgments
We thank Dong Lai and Jim Fuller for helpful discussions during the conception and the development of this study. 
This work was supported by NSF AST-2054353.
H.Y. acknowledges the support of the Sherman Fairchild Foundation.

% \clearpage
\appendix

\section{Amplitude of \lowercase{$m\neq 0$}   daughters}
\label{sec:q_b}
In Eqs.~(\ref{eq:q_b_ss_appx}) and (\ref{eq:q_c_ss_soln}) we assumed the daughter modes' amplitudes are given by their steady-state values. While this is a good approximation for `mode $c$' (daughters with $m_c=0$), as we explain here the problem may be more involved for `mode $b$' (daughters with $l_b=-m_b=4$). 

The equation governing such a mode $b$ is given by [see Eq.~(\ref{eq:ode_mode_amp_no_Uab})]
\begin{eqnarray}
    \dot{q}_b + (i \omega_b &+& \gamma_b) q_b 
	=  i\omega_b \kappa_b (q_a^\ast q_a^\ast - 2q_a^\ast U_a^\ast),
\end{eqnarray}
where,  as explained in Section~\ref{sec:nl_freq_damp} (also see discussion below), we can ignore the linear tidal forcing on mode $b$, i.e., the $U_b$ term. We can decompose the parent mode (mode $a$) as 
\begin{equation}
    q_a = q_{a,{\rm dyn}} + q_{a, {\rm eq}} =  q_{a,{\rm dyn}} + U_a,
\end{equation}
where $q_{a, {\rm eq}}\equiv U_a$ is the equilibrium tide solution of mode $a$ [which can be obtained from Eq.~(\ref{eq:ode_mode_amp_general}) when we ignore the nonlinear couplings and treat $|\dot{q}_a|, |\gamma_a q_a| \ll |\omega_a q_a|$].  We thus have
\begin{equation}
    \dot{q}_b + (i \omega_b + \gamma_b) q_b 
	=  i\omega_b \kappa_b (q_{a, {\rm dyn}}^\ast q_{a, {\rm dyn}}^\ast + U_a^\ast U_a^\ast).
	\label{eq:qb_eom}
\end{equation}

In the main text, we focused on the steady-state solution of $q_b$ driven by a free-oscillating parent. That is, we assumed $q_a$ contains only the dynamical component $q_{a,{\rm dyn}}$ which oscillates at a single frequency $\omega_a$, and found [see Eq.~(\ref{eq:q_b_ss_soln})]
\begin{equation}
    q_{b,{\rm ss}} = \frac{\omega_b \kappa_b}{\Delta_b - i\gamma_b} \left(q_{a,{\rm dyn}}^\ast\right)^2,
    \label{eq:q_b_ss_appx}
\end{equation}
where $\Delta_b  = (\omega_b + 2\omega_a)$. Note, however, that the steady state solution $q_{b,\rm ss}$ neglects the $U_a^\ast U_a^\ast$ term in Eq.~\ref{eq:qb_eom} and it  neglects the `transient' part of the solution for $q_b$.  We will refer to the latter as the history term since it depends on the past history of $q_b(t)$ from previous pericenter passages.  Here we will show that the $U_a^\ast U_a^\ast$ term  should always be insignificant but not necessarily the history term.

We will make two simplifications in our analysis. First, we do not  explicitly solve for the instantaneous value of $q_b$ in the vicinity of a pericenter passage for simplicity. 
As we will see, this does not preclude us from obtaining 
a qualitative
estimate of the history term  due to previous pericenter passages.  Second, we treat the parent mode as if it is unperturbed by  nonlinear interactions. As a result, we assume that the dynamical component of the parent mode, $q_{a, \rm dyn}$, when far away from the pericenter, oscillates at $\omega_a$  and not at the nonlinearly shifted frequency (cf. Sec.~\ref{sec:nl_freq_damp}).

\subsection{The $U_a^\ast U_a^\ast$ term}
First we consider the drive due to the $U_a^\ast U_a^\ast$ term. Similar to the one-kick amplitude of the parent $\Delta q_{a, 1}$, we can define a one-kick amplitude of mode $b$ at each pericenter passage due to the $U_a^\ast U_a^\ast$ term as 
\begin{equation}
    \left(\Delta q_{b,1}\right)_{U_a^2} = \int i \omega_b \kappa_b W_{l_am_a}^2Q_a^2\left(\frac{M_\ast}{M}\right)^2\left(\frac{R}{D}\right)^{l_b+2}e^{i\left[(\omega_b + m_b \Omega_{\rm s})\tau - m_b \Phi\right]},
\end{equation}
where we used the fact that $l_b = 2l_a$ and $m_b = -2m_a$. 
If we define 
\begin{equation}
        K'_{lm}(\omega) = \frac{\omega_0} {2\pi}\int \left[\frac{D_{\rm peri}}{D(\tau)}\right]^{l+1} e^{i\left[\omega \tau - m\Phi(\tau)\right]} d\tau,
\end{equation}
as a modified temporal overlap, 
then we can write the one-kick amplitude of the daughter mode as 
\begin{equation}
    \left(\Delta q_{b,1}\right)_{U_a^2} = i 2\pi \kappa_b W_{22}^2 Q_a^2\left(\frac{\omega_b}{\omega_0}\right) \left(\frac{M_\ast}{M}\right)^2\left(\frac{R}{D_{\rm peri}}\right)^{6}K'_{5, -4}(\omega_b-4\Omega_{\rm s}),
\end{equation}
where we have plugged in $l_a=m_a=2$ for the parent and $l_b=-m_b=4$ for the daughter. 
For typically values ($a_{\rm orb}{=}1\,{\rm AU}$, $e_{\rm orb}{=}0.98$, $\omega_a{=}5.6\times 10^{-4}\,{\rm rad/s}$, and $\omega_b {=} -1.9 \omega_a$), we find  $\big{|}\left(\Delta q_{b}\right)_{U_a^2}\big{|}\simeq 0.4 |\Delta q_{a, 1}^2|$. 

Although intially $\left(\Delta q_{b,1}\right)_{U_a^2}$ is comparable to the steady-state solution $q_{b,{\rm ss}}$  [Eq.~(\ref{eq:q_b_ss_appx})], as the system starts to grow diffusively, its effect will soon become subdominant. This can be seen by noticing that even if $\left(\Delta q_{b,1}\right)_{U_a^2}$ can grow diffusively itself (e.g., due to a random $P_{\rm orb}$), after $k$ pericenter passages, it only increase the amplitude of $q_b$ by $\sqrt{k} |\left(\Delta q_{b,1}\right)_{U_a^2}|\propto k^{1/2}$ on average. On the other hand, $q_{b, {\rm ss}}\sim k|q_{a,{\rm dyn}}|^2 \propto k$ because each $|q_{a, {\rm dyn}}|$ grows as $\sqrt{k}$. Consequently, the significance of the $q_{b, {\rm ss}}$ term increases as $\sqrt{k}$. In fact, the dominance of $q_{b, {\rm ss}}$ is further enhanced by the $\omega_b/\Delta_b$ factor, especially for the most resonant daughter mode with the smallest $|2\omega_a + \omega_b|$. 

We therefore conclude that the modification to $q_b$ due to the $\left(U_a^\ast\right)^2$ term can be ignored. It is also worth noting that the drive from $\left(U_a^\ast\right)^2$ is stronger than the direct tidal force on mode $b$, $U_b\propto (M_\ast/M)(R/D_{\rm peri})^{l_b+1}$, by a factor of $(M_\ast/M)(R/D_{\rm peri})\left(\kappa_b W_{22}^2Q_a^2/W_{44}Q_b\right)\simeq 6\left(\kappa_b W_{22}^2Q_a^2/W_{44}Q_b\right)$. It thus justifies why we can also ignore the daughter modes' linear coupling to the tide.

\subsection{The $q_{a,{\rm dyn}}^\ast q_{a,{\rm dyn}}^\ast$ term}

We now consider the effect of the history term on the daughter.  

If we define $c_b = q_b\exp(-2i\omega_a t)$ then by Eq.~(\ref{eq:qb_eom}), 
\begin{equation}
    \dot{c}_b+\left(i\Delta_b + \gamma_b\right)c_b = i\omega_b V_b,
\end{equation}
where $V_b \equiv \kappa_b q_a^\ast q_a^\ast \exp(-2i\omega_a t)$. Our definition of $V_b$ does  not include the equilibrium tide contribution $U_a^\ast U_a^\ast$ and $U_b$ since we showed above that they are insignificant.  For the same reason, here and below we drop the ``dyn'' subscript on the parent.  Note that if we ignore the parent's nonlinear frequency corrections, then away from pericenter $q_a \sim e^{-i\omega_a t}$ and thus $V_b$ is a constant. Near pericenter, however, $q_a$ has an additional time-dependence due to the kick $\Delta q_{a, 1}$ the parent receives  over a timescale $1/\Omega_{\rm peri}$.  As we will see, it is this effect that constitutes the history term we are interested in. 

The general solution for $c_b$ is given by
\begin{eqnarray}
    c_b(t) 
    &=& e^{-(i\Delta_b + \gamma_b)t}\int_{t_0}^t i \omega_b V_b(\tau) e^{(i\Delta_b + \gamma_b)\tau}d\tau. \nonumber \\
    &=&\frac{\omega_b}{\Delta_b - i\gamma}V_b\Big{|}_{t_0}^t - e^{-(i\Delta_b+\gamma_b)t}\omega_b\int^t_{t_0} \frac{\dot{V}_b(\tau)}{\Delta_b -i\gamma_b} e^{(i\Delta_b + \gamma_b)\tau}d\tau,
    \label{eq:c_b_soln}
\end{eqnarray}
where the initial time is $t_0$ and we performed integration by parts to get the second line. For future convenience we set $V_b(t_0)=0$ and thus drop the initial condition. Note that the first term in Eq.~(\ref{eq:c_b_soln}) recovers the steady-state solution, Eq.~(\ref{eq:q_b_ss_soln}), and it depends only on the instantaneous value of $q_a$. The second term, on the other hand, captures the past history. For a free oscillator, $q_a \sim e^{-i\omega_a t}$ and  $\dot{V}_b=0$, and thus the value of $q_b$ is independent of the past history. 

When the system is coupled to the tide, however, we have $\dot{V}_b\sim \Omega_{\rm peri} V_b$ in the vicinity of  the pericenter. 
First consider a mode $b$ for which $|\Delta_b|\gg\Omega_{\rm peri}\gg \gamma_b$.  These inequalities hold for all the daughters in our mode  networks with the exception of the $l_b=-m_b=4$, $\omega_b<0$ f-mode, which we consider separately below. For large detuning, if we keep performing integration by parts, we get 
\begin{equation}
    c_b(t)\simeq
    \frac{\omega_b}{\Delta_b}V_b(t) 
    + i \frac{\omega_b}{\Delta_b}\frac{\dot{V}_b(t)}{\Delta_b} 
    + ie^{-i\Delta_b t}\omega_b\int^t \frac{\ddot{V}_b(\tau) }{\Delta_b^2 }e^{i\Delta_b \tau}d\tau=...,
\end{equation}
where we dropped $\gamma_b$ to simplifiy the notation. Note that after the $n$'th iteration of integration by parts, we have a correction that depends only on the instantaneous value $\propto V_b^{(n-1)}/\Delta_b^{n-1}$, and an integrand  $\propto V_b^{(n)}/\Delta_b^n\sim (\Omega_{\rm peri}/\Delta_b)^n V_b$, where $V_b^{(n)}$ is the $n$'th time derivative of $V_b$. Since $|\Delta_b| \gg \Omega_{\rm peri}$, the history-dependent term gets progressively smaller with increasing $n$, and the instantaneous corrections to Eq.~(\ref{eq:q_b_ss_soln}) form a converging series (in fact, the corrections are non-zero only around a pericenter passage). 
This is analogous to the fact that the linear tide can be well approximated by its instantaneous equilibrium component when the tidal forcing frequency is much smaller than the mode frequency.

For the most resonant daughter mode $b$ (i.e., the $l_b=-m_b=4$, $\omega_b<0$ f-mode), it is possible to have $|\Delta_b| < \Omega_{\rm peri}$. In this case, the series expansion formed by integration by parts does not converge. Instead, we need to directly solve  Eq.~(\ref{eq:c_b_soln}).  To do so, we consider the following simple model of $V_b$ near the $k$'th pericenter passage (corresponding to time $t=t_k$) \begin{equation}
V_b(t) = 
\left\{
\begin{array}{@{}ll@{}}
-\Delta V_{bk},&\text{if $t< t_{k}-\dfrac{\pi}{\eta\Omega_{\rm p}}$}, \\
\Delta V_{bk}\sin \left[\eta\Omega_{\rm p} (t-t_k)\right],&\text{if $ t_{k}-\dfrac{\pi}{\eta\Omega_{\rm p}}\leq t\leq t_{k} + \dfrac{\pi}{\eta\Omega_{\rm p}}$},\\
\Delta V_{bk},&\text{if $t> t_{k}+\dfrac{\pi}{\eta\Omega_{\rm p}}$},
\end{array}\right.
\end{equation}
where $\eta \sim 1$ is a correction on the characteristic timescale over which $q_a$ changes, and we rewrote $\Omega_{\rm peri}$ as $\Omega_{\rm p}$ in order to reduce notational clutter. 
With this simple model we can easily evaluate the integration around $t_k$ as
\begin{eqnarray}
&&\int^{t_k+\pi/\eta\Omega_{\rm p}}_{t_k-\pi/\eta\Omega_{\rm p}} \dot{V}_b(\tau) e^{(i\Delta_b+\gamma_b) \tau}d\tau\nonumber \\
\simeq&& \left[\frac{\eta\Omega_{\rm p}}{(\Delta_b + \eta\Omega_{\rm p} - i \gamma_b)}\sin\left(\frac{\Delta_b + \eta\Omega_{\rm p} }{\eta\Omega_{\rm p} }\pi\right)  
+ \frac{\eta\Omega_{\rm p}}{(\Delta_b - \eta\Omega_{\rm p} - i \gamma_b)}\sin\left(\frac{\Delta_b - \eta\Omega_{\rm p} }{\eta\Omega_{\rm p} }\pi\right) 
\right]\Delta V_{bk},
\end{eqnarray}
where we ignored dissipation over a time $2\pi/\eta\Omega_{\rm p}$. We can now write mode $b$'s amplitude as\footnote{In fact, this solution applies to the case where $\Omega_{\rm peri}<|\Delta_b|$ as well. It is just less apparent to see why in the large $\Delta_b$ case the amplitude of mode $b$ is independent of its history from Eq.~(\ref{eq:c_b_explicit_soln}) than from the series expansion formed by consecutive integration by part.} 
\begin{eqnarray} 
c_b(t) &\simeq \dfrac{\omega_b}{\Delta_b - i\gamma_b} 
\left\{ 
V_b(t) + e^{-(i\Delta_b+\gamma_b)t} 
    \left[     \dfrac{\eta\Omega_{\rm p}\sin\left(\dfrac{\Delta_b \pm \eta\Omega_{\rm p} }{\eta\Omega_{\rm p} }\pi\right)}{(\Delta_b \pm \eta\Omega_{\rm p} - i \gamma_b)}  
	\right]
   \mathlarger{\sum}\limits_{k}^{t_k<t}
	\Delta V_{bk} 
\right\}.\label{eq:c_b_explicit_soln}
\end{eqnarray}
Therefore, in addition to the instantaneous term $V_b(t)$, there in principle should also be a history-dependent term $\sum \Delta V_{bk}$. Physically, this case can be understood by the following. As the parent mode's amplitude changes at each pericenter passage over a timescale $1/\eta\Omega_{\rm peri}$, its frequency content is broadened from a single delta function at $\omega_a$ to a band covering $\omega_a \pm \eta\Omega_{\rm peri}/2$. Since the parent-daughter detuning is small, $|\Delta_b|<\Omega_{\rm peri}$, the broadened drive from the parent can now resonantly excite the daughter. This thus gives the daughter a ``dynamical'' component that depends on the past history ($\sum \Delta V_{bk}$). By contrast, when the daughter is not resonant the instantaneous ``equilibrium'' component dominates.

If $\sum \Delta V_{bk}$  grows diffusively (since $\Delta V_{bk}\propto q_a^\ast \Delta q_{a,1}$, and the phase of $q_a^\ast$ can be random), then both $V_b$ and $\sum \Delta V_{bk}$ grow with   the number of pericenter passages $k$ as $\propto k$. Therefore, the history term due to the kick on $q_a$, unlike the one due to $U_a$, can be potentially important. For simplicity, we drop it in the analysis of this paper and defer its consideration to   future work.


\bibliography{ref}{}
\bibliographystyle{aasjournal}



\end{document}